\documentclass[twocolumn,aps,pra,groupedaddress,superscriptaddress,amsmath,amssymb,showpacs,noeprint]{revtex4-2}

\usepackage[T1]{fontenc}
\usepackage{graphicx}
\graphicspath{{./Figures/}}
\usepackage{dcolumn}
\usepackage{bm}
\usepackage[dvipsnames]{xcolor}
\usepackage{hyperref}
\hypersetup{colorlinks,linkcolor={MidnightBlue},citecolor={MidnightBlue},urlcolor={MidnightBlue}}  

\usepackage{amsmath}
\usepackage{graphicx}
\usepackage{dcolumn}
\usepackage{bm}
\usepackage[dvipsnames]{xcolor}
\usepackage{hyperref}
\hypersetup{colorlinks,linkcolor={MidnightBlue},citecolor={MidnightBlue},urlcolor={MidnightBlue}}  

\usepackage{physics}
\usepackage{amsfonts}
\usepackage{verbatim}
\usepackage{amsmath}
\usepackage{csquotes}
\usepackage{color}
\usepackage{soul}

\usepackage{float}

\usepackage{physics}
\usepackage{amsfonts}
\usepackage{verbatim}
\usepackage{amsmath}
\usepackage{csquotes}
\usepackage{color}
\usepackage{soul}
\usepackage{amsthm}
\usepackage{bm}
\usepackage{graphicx}
\usepackage[normalem]{ulem}

\newcommand{\prt}[1]{\left(#1\right)}

\newcommand{\prtq}[1]{\left[#1\right]}

\newcommand{\prtg}[1]{\left\{#1\right\}}

\newcommand{\Id}{\mathbb{I}}

\newcommand{\ad}{a^\dagger}

\newcommand{\add}{a^{\dagger 2}}
\newcommand{\adn}{\prt{a^\dagger}^n}
\newcommand{\an}{a^n}

\newcommand{\hc}{\textrm{h.c.}}
\newcommand{\cst}{\rho_A^\textup{st}}

\newcommand{\rth}{\rho_A^\textup{th}}
\newcommand{\intH}{H_\textup{int}}
\newcommand{\alphadue}{\abs{\alpha}^2}
\newcommand{\alfsn}{\prt{\alpha^*}^n}

\newcommand{\Lb}{\mathcal{L}}
\newcommand{\D}{\mathcal{D}}

\newcommand{\K}{\mathcal{K}}
\newcommand{\J}{\mathcal{J}}

\newcommand{\R}{\mathit{R}}

\newcommand{\nuno}{n_1}
\newcommand{\ndue}{n_2}

\newcommand{\gamloc}{\gamma_{\textup{loc}}}
\newcommand{\Jcorr}{J_{\textup{corr}}}
\newcommand{\nwt}{\bar{n}_{\omega,T}}
\newcommand{\ndwt}{\bar{n}_{2\omega,T}}
\newcommand{\nlwt}{\bar{n}_{l\omega,T}}

\newcommand{\prtT}[1]{(#1)}

\newcommand{\prtgT}[1]{\{#1\}}

\newcommand{\phidot}{\dot \phi}
\newcommand{\varphidot}{\dot \varphi}
\newcommand{\Cos}[1]{\cos\left( #1\right)}
\newcommand{\Sin}[1]{\sin\left( #1\right)}
\newcommand{\eff}{{\rm eff}}
\newcommand{\alphaN}{\eta}

\begin{document}

\title{Two-photon-interaction effects in the bad-cavity limit}

\author{Nicol\`o Piccione}
\email{nicolo.piccione@univ-fcomte.fr}
\affiliation{Institut UTINAM, CNRS UMR 6213, Universit\'{e} Bourgogne Franche-Comt\'{e}, Observatoire des Sciences de l'Univers THETA, 41 bis avenue de l'Observatoire, F-25010 Besan\c{c}on, France}

\author{Simone Felicetti}

\affiliation{Istituto di Fotonica e Nanotecnologie, Consiglio Nazionale delle Ricerche, Piazza Leonardo da Vinci 32, I-20133 Milano, Italy}

\affiliation{Universit\'{e} de Paris, Laboratoire Mat\'{e}riaux et Ph\'{e}nom\`{e}nes Quantiques, Centre National de la Recherche Scientifique, F-75013 Paris, France}

\author{Bruno Bellomo}
\affiliation{Institut UTINAM, CNRS UMR 6213, Universit\'{e} Bourgogne Franche-Comt\'{e}, Observatoire des Sciences de l'Univers THETA, 41 bis avenue de l'Observatoire, F-25010 Besan\c{c}on, France}

\begin{abstract}

Various experimental platforms have proven to be valid testbeds for the implementation of nondipolar light-matter interactions, where atomic systems and confined modes interact via two-photon couplings. Here, we study a damped quantum harmonic oscillator interacting with $N$ two-level systems via a two-photon coupling in the so-called bad-cavity limit, in the presence of finite-temperature baths and coherent and incoherent drivings.
We have succeeded in applying a recently developed adiabatic elimination technique to derive an effective master equation for the two-level systems, presenting two fundamental differences  compared to the case of a dipolar interaction: an enhancement of the two-level systems spontaneous-like emission rate, including a thermal contribution and a quadratic term in the coherent driving, and an increment of the effective temperature perceived by the two-level systems. These differences give rise to striking effects in the two-level systems dynamics, including a faster generation of steady-state coherence and a richer dependence on temperature of the collective effects, which can be made stronger at higher temperature.  

\end{abstract}

\maketitle

\section{Introduction}

Atomic systems interacting with confined photonic or phononic modes represent one of the most studied classes of quantum-optical systems. On the one hand, the confinement may induce modifications of single atom absorption and emission rates such as the well-known Purcell effect \cite{BookHaroche2006}. On the other hand, the collective nature of such interactions gives rise to a rich quantum phenomenology characterized, for example, by the emergence of quantum phase transitions~\cite{KirtonReview} and by the qualitative modifications of optical properties~\cite{benedict2018super}.
Concerning the latter, a sub and a superradiant regime have been identified, respectively characterized by the dampening or the amplification of atomic absorption and emission rates  with respect to the independent-emitter case \cite{GROSS1982301}. These regimes have been  extensively studied also in the presence of coherent or incoherent optical drivings~\cite{Meiser2009,Meiser2010I,Meiser2010II,Auffeves2011,PhysRevA.96.023863,kirton2018,PhysRevA.98.063815,PhysRevA.99.033845}. 
Much attention has been devoted to the so-called bad-cavity limit in which the confined mode is strongly dampened with respect to the interaction with the atoms \cite{Meiser2009,Meiser2010I,Meiser2010II,Auffeves2011,kirton2018,PhysRevA.98.063815}. In this context, the effective dynamics of the atoms can be obtained by adiabatically eliminating the confined mode \cite{Bonifacio1971(I),Bonifacio1971(II),Azouit2017,ThesisAzouit2017}.

Besides the fundamental interest, collective quantum phenomena induced by light-matter interactions can be exploited in a variety of applications. In particular, the sub and superradiant regimes may be associated to the generation of collective states of the emitters, which are of great interest for quantum sensing~\cite{PhysRevLett.104.073602,toth2014quantum},  generation of nonclassical states~\cite{jahnke2016giant}, photon storage~\cite{PhysRevX.7.031024}, and excitation transfer~\cite{garcia2017long}.
This phenomenology is of high experimental relevance, as collective light-matter interactions can be controllably implemented in a broad range of atomic and solid-state quantum systems, such as cold atoms~\cite{PhysRevLett.115.063601}, trapped ions~\cite{PhysRevLett.76.2049}, metamaterials~\cite{PhysRevLett.119.053901}, plasmonic cavities~\cite{PhysRevLett.102.077401}, color centres in diamonds~\cite{angerer2018superradiant}, quantum dots~\cite{scheibner2007superradiance}, and superconducting circuits~\cite{mlynek2014observation}.

To the best of our knowledge, collective radiative phenomena have not so far been analyzed for two-photon (2ph) interactions. However, it has been recently predicted that using atomic or solid-state systems it is possible to implement nondipolar light-matter couplings, where the linear interaction is inhibited and where quantum emitters and localized bosonic modes interact via the exchange of two excitation quanta. In particular, such two-photon couplings can be observed by engineering superconducting atom-resonator 
systems~\cite{felicetti_two-photon_2018,PhysRevA.98.053859} or by applying analog quantum simulation schemes in trapped-ions~\cite{felicetti_spectral_2015,PhysRevA.97.023624,PhysRevA.99.032303} or ultracold atoms~\cite{Schneeweiss_2018, Dareau_2018}.
Notice that  nondipolar transitions have already been observed using superconducting artificial atoms~\cite{PhysRevLett.121.060503}, and that quantum-simulation techniques have already been experimentally applied to observe the physics of fundamental dipolar light-matter interaction models in extreme regimes of parameters~\cite{Dareau_2018, PhysRevX.8.021027}.
On the dissipative side, two-photon relaxation~\cite{Minganti2016,Malek2019} and pumping~\cite{Minganti2016} have also been theoretically analyzed and experimentally implemented~\cite{Leghtas2015}.
The fast-growing interest in two-photon couplings is motivated by a rich phenomenology, characterized by counter-intuitive spectral  features~\cite{PhysRevA.85.043805,duan2016two,maciejewski2017novel,PhysRevA.99.013809,PhysRevA.99.013815,Rico2020}, high-order quantum optical nonlinearities~\cite{felicetti_two-photon_2018,PhysRevA.98.053859,zou2019multiphoton}, and quantum phase transitions~\cite{garbe_superradiant_2017,PhysRevA.97.053821,PhysRevA.100.033608,garbe2019dissipation,Cui2020}. In turn, this phenomenology can be exploited in different quantum-information applications~\cite{PhysRevLett.122.123604,casanova2018connecting,PhysRevA.99.023854}. We finally stress that the two-photon coupling analyzed here differs from other physical situations for which the term \enquote{two-photon} is used. Some examples are: two-photon excitations (see chapter 6.7 of Ref.~\cite{Book_Loudon2000}), two-photon absorption~\cite{Rumi2010Two-Photon}, two-plasmon emission~\cite{Rivera2016Shrinking}, and two-photon emission coming from strong light-matter coupling~\cite{Flick2018Strong}.

In this Letter, we study the dynamics of a damped harmonic oscillator (HO) interacting with an ensemble of two-level systems (TLSs) in the bad-cavity limit in the case of a two-photon coupling. By applying a recently developed approach to perform adiabatic elimination in open quantum systems~\cite{ThesisAzouit2017,Azouit2017}, we derive an effective master equation for the TLSs that takes into account the coupling with finite-temperature baths as well as coherent and incoherent optical drivings. Our analytical and numerical analysis of the time evolution and steady-state behavior unveils a unexpected collective phenomenology induced by nondipolar light-matter interactions.
Compared to the dipolar case, the two-photon coupling introduces the possibility to enhance  the absorption and emission processes, and leads to a higher resilience of sub- and superradiance with respect to the baths temperature.

\section{Physical models}

We study a system composed of a damped HO interacting via a resonant Jaynes-Cummings Hamiltonian with $N$ TLSs in the bad-cavity limit~\cite{Bonifacio1971(I),Bonifacio1971(II),ThesisAzouit2017}, comparing the one-photon (1ph) and 2ph interaction cases.
The two models are described by  the  Hamiltonians
\begin{equation}
H_l = \hbar\omega \ad a+ \frac{l \hbar \omega}{2}J_z +\hbar g\prtq{a^l J_+ + \prt{\ad}^l J_-},
\end{equation}
where $l=1$ for the 1ph case and $l=2$ for the 2ph one,  $\omega$ is the frequency  of the HO and $l\omega$ the one of the TLSs (i.e., we consider a resonant interaction in both cases), $g$ is the coupling parameter  between the HO and the TLSs, $a$  and $\ad$ are the usual annihilation and creation operators of a HO, while  $J_z = \sum_{i=1}^{N} \sigma_z^{(i)}$ and $J_\pm = \sum_{i=1}^{N} \sigma_\pm^{(i)}$, where $\sigma_z$, $\sigma_-$, and $\sigma_+$ are, respectively,  the $z$-Pauli,  the lowering, and the raising operators of a TLS. The ground and the excited energy levels of each TLS are indicated, respectively, by $\ket{g}$ and $\ket{e}$.
In Appendix~\ref{APPSec:CircuitModel} we provide an example of a possible implementation with superconducting circuits~\cite{blais2020circuit} of the above Hamiltonian for the case $l=2$, by generalizing the study done in Ref.~\cite{PhysRevA.98.053859} to the case of more than one TLS.

We suppose that the HO and each TLS are each in contact with an independent thermal bath at temperature $T$ (equal for all baths) and that a resonant coherent pumping on the HO and an incoherent local pumping on the TLSs are available.
In the interaction picture, using a phenomenological approach \cite{BookBreuer2007,Rivas2010,Giorgi2020}, the master equation for the global density matrix $\rho_G$ is
\begin{equation}\label{eq:FullModel}
\dot{\rho}_G= - i g \comm{\an  J_++ \adn J_-}{\rho_G} + \Lb_{\textup{HO}}(\rho_G)+\Lb_Q (\rho_G),
\end{equation}
where $\Lb_{\textup{HO}} (\bullet)$ and $\Lb_Q (\bullet)$ are dissipators acting, respectively, on the HO and on the TLSs, given by 
\begin{align}
	\Lb_{\textup{HO}} (\bullet)= &   -i \comm{\prt{\beta^* a+\beta \ad}}{\bullet} \nonumber\\
	&+k\prtq{\prt{1+\nwt}\D_{a}\prt{\bullet}+\nwt\D_{\ad}\prt{\bullet}}, \nonumber\\
	\Lb_Q (\bullet)=&  \sum_{i=1}^{N} \left[\gamloc\prt{1+\nlwt }\D_{\sigma_-^{(i)}}\prt{\bullet} \right. \nonumber \\
	&\left.+\prt{\gamloc\nlwt+P}\D_{\sigma_+^{(i)}}\prt{\bullet}\right],	
	\end{align}
where $\D_{X}  (\bullet)= X \bullet  X^\dagger -\frac{1}{2}\{X^\dagger X,\bullet\}$, $k$ and $\gamloc$ are the relaxation rates of, respectively, the HO and each TLS due to the local couplings with their own thermal baths ($\gamloc$ is assumed to be the same for all the TLSs),  
$\beta$  characterizes the interaction between the HO and the coherent field, $P$ quantifies the action of the incoherent pumping on each TLS, and $\nwt= [e^{\hbar \omega/(k_B T)}-1]^{-1}$, $k_B$ being  the Boltzmann constant. The coherent pumping is treated in the rotating-wave approximation, being  $\abs{\beta} \ll \omega$.
The phenomenological approach is justified because we consider the TLSs and the HO weakly coupled  ($g  \ll \omega$) \cite{BookBreuer2007},  the HO weakly coupled to its bath ($k \ll \omega$) \cite{BookBreuer2007}, and the external coherent field resonant with the HO \cite{Rivas2010}.

\section{Adiabatic elimination}

By applying a recently introduced adiabatic elimination technique~\cite{ThesisAzouit2017,Azouit2017}   we have been able to derive an effective master equation for the reduced density matrix of the TLSs, $\rho=\mathrm{Tr}_{\textup{HO}}\{\rho_G\}$ (see Appendices~\ref{APPSec:AdiabaticEliminationGeneral} and~\ref{APPSec:AdiabaticEliminationApplied} for a review of this technique, the detailed derivation, and some comments on the validity range of the adiabatic elimination):
\begin{align}
\label{eq:TheReducedMasterEquation}
\dot{\rho} 
&= -i g \comm{\alpha^l J_+ + \prt{\alpha^*}^l J_-}{\rho} + \Lb_Q(\rho)\nonumber \\ 
&\quad +\gamma_l\prtq{n_l \D_{J_+}\prt{\rho}+ \prt{1+n_l} \D_{J_-}\prt{\rho}},
\end{align}
where we recall that $l=1$ for the 1ph case and $l=2$ for the 2ph one, and
\begin{equation}
\begin{gathered}\label{eq:MEparameters}
\alpha=-\frac{2 i \beta}{k}, \quad \gamma_1=\frac{4g^2}{k}, \quad  \nuno=\bar{n}_{\omega,T},\\
\gamma_2 = \gamma_1 \prt{1 + 2 \nuno + 4 \alphadue}, \quad
\ndue = \nuno \frac{\nuno + 4\alphadue}{1+2\nuno + 4\alphadue}.
\end{gathered}
\end{equation}

As expected,  even in the 2ph case the adiabatic elimination gives rise  to collective dissipative terms [second line of Eq.~\eqref{eq:TheReducedMasterEquation}]. We observe that differently from the case of collective radiative phenomena induced by the interaction of different atoms with a common vacuum field \cite{GROSS1982301}, here the collective phenomena result from the coupling with a common damped HO.
Notice that, although Eq.~\eqref{eq:TheReducedMasterEquation} retains its formal structure when changing $l$ (see Appendix~\ref{APPSec:MathematicalMapping} for details), the effective parameters $\alpha^l, \gamma_l$, and $n_l$ coming from the adiabatic elimination depend differently in the two models on the physical parameters  $g$, $\beta$, $k$, $\omega$, and  $T$ [see Eq.~\eqref{eq:MEparameters}]. This  results in profound physical differences between the 1ph case and the 2ph one, leading to unexpected effects specific to the 2ph case.
In particular, we can identify three main modifications.
A first evident difference regards the dependence  of the unitary driving term on $\alpha$, which is linear in the 1ph case and quadratic in the 2ph one.
An even more striking difference concerns the collective relaxation rate $\gamma_l$ which, only in the 2ph case, depends on the parameters characterizing the state of the HO at order zero,  $n_1$ and $\alpha$ (see Appendix~\ref{APPSec:AdiabaticEliminationApplied}). Finally, the coherent pumping increases the temperature of the effective collective bath seen by the TLSs, generated by the adiabatic elimination of the HO.
In particular, setting $\ndue =\bar{n}_{2\omega,T^*}= [e^{2\hbar\omega/\prt{k_B T^*}} - 1]^{-1}$,
the temperature of this collective bath is
\begin{equation}
T^*= \frac{2 \hbar \omega}{k_B}\prtq{\ln \prt{\frac{e^{2 \hbar\omega/(k_B T)} - 2}{1+ 4 \alphadue\prt{e^{\hbar \omega/(k_B T)} - 1}} + 2}}^{-1}.
\end{equation}
Notice that when $\alpha=0$ the temperature of this collective bath would be the same as that of the original bath of the HO ($T^*=T$).
The peculiar form of $\gamma_2 $ and $n_2$,  especially their quadratic dependence on $|\alpha|$, can be useful to manipulate the dynamics of the TLSs, possibly 
enhancing  their absorption and emission processes.

In the following, we discuss  the physical consequences of these differences. In order to check the validity of the adiabatic elimination, we will show in several figures numerical simulations of the full model of Eq.~\eqref{eq:FullModel}.

\begin{figure}[t!]
\centering	
\includegraphics[width=0.48\textwidth]{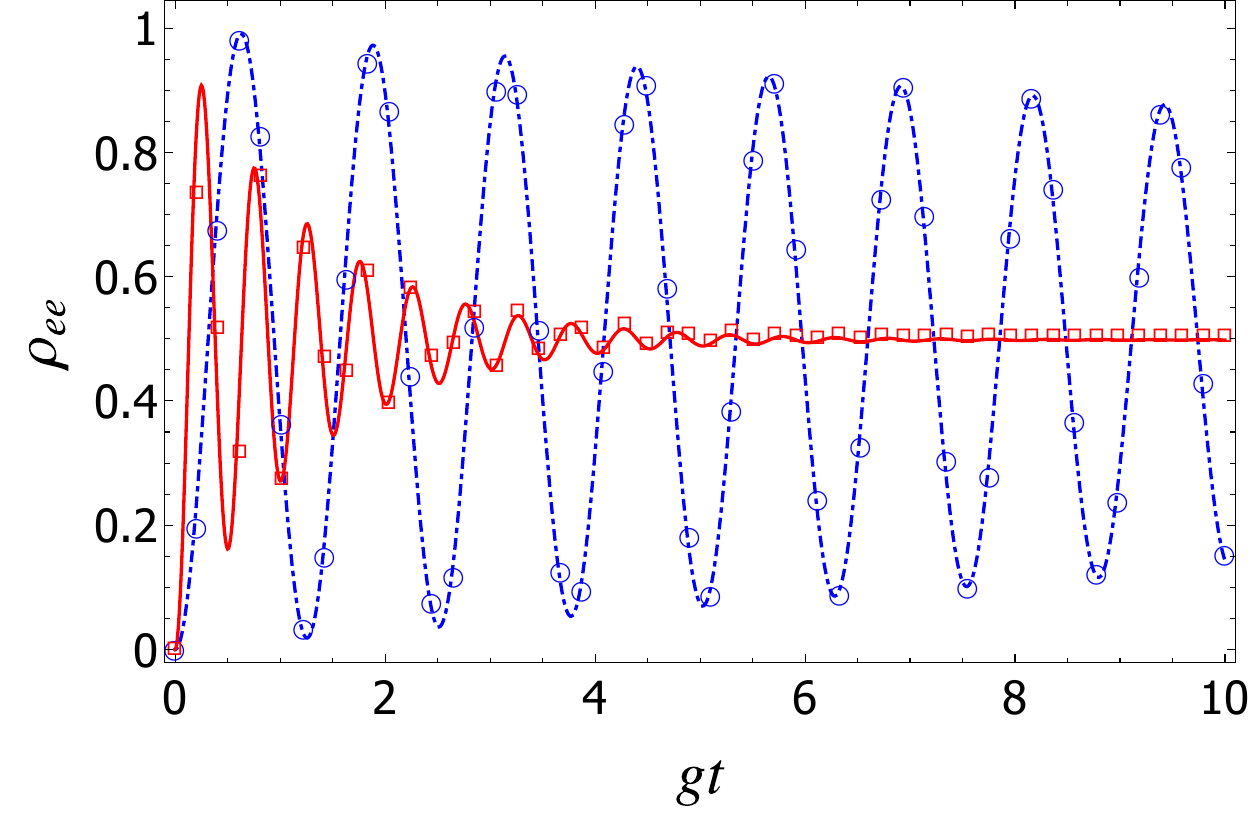}
\caption{Time evolution of the excited-state population of one TLS, $\rho_{ee}$, with physical parameters $\beta= 1.25k$ (so that $\abs{\alpha}=2.5$), $\gamloc=0$, $g=0.01k$, $T=0$, and $P=0$. The dot-dashed blue line and the continuous red line are the curves obtained by using the effective model of Eq.~\eqref{eq:TheReducedMasterEquation} for, respectively, the 1ph and 2ph models. Empty markers show discrete points obtained from the numerical simulation of the full model of Eq.~\eqref{eq:FullModel}.
In the 2ph model the steady state is clearly reached much faster.}
\label{fig:EnhancedPurcellLikeEffect}
\end{figure}

\section{Coherent driving effects: faster dynamics and robust steady-state coherence}

In order to focus on the effects due to the coherent pumping on the HO, let us consider the case of zero temperature and no local incoherent pumping on the TLSs.
For $T=0$ and $P=0$, Eq.~\eqref{eq:TheReducedMasterEquation} simplifies and
 $\gamma_2 = \gamma_1 (1 + 4 \alphadue)$.

The quadratic dependence of $\gamma_2$ on $|\alpha|$ can be exploited to make the system reach much faster its  steady state in the 2ph case. This is shown in  Fig.~\ref{fig:EnhancedPurcellLikeEffect},  comparing the dynamics of one TLS (henceforth we use the notation $\langle x | \rho|  y \rangle =\rho_{xy} $) for the two models.
 
\begin{figure} [t!]
\centering	\includegraphics[width=0.48 \textwidth]{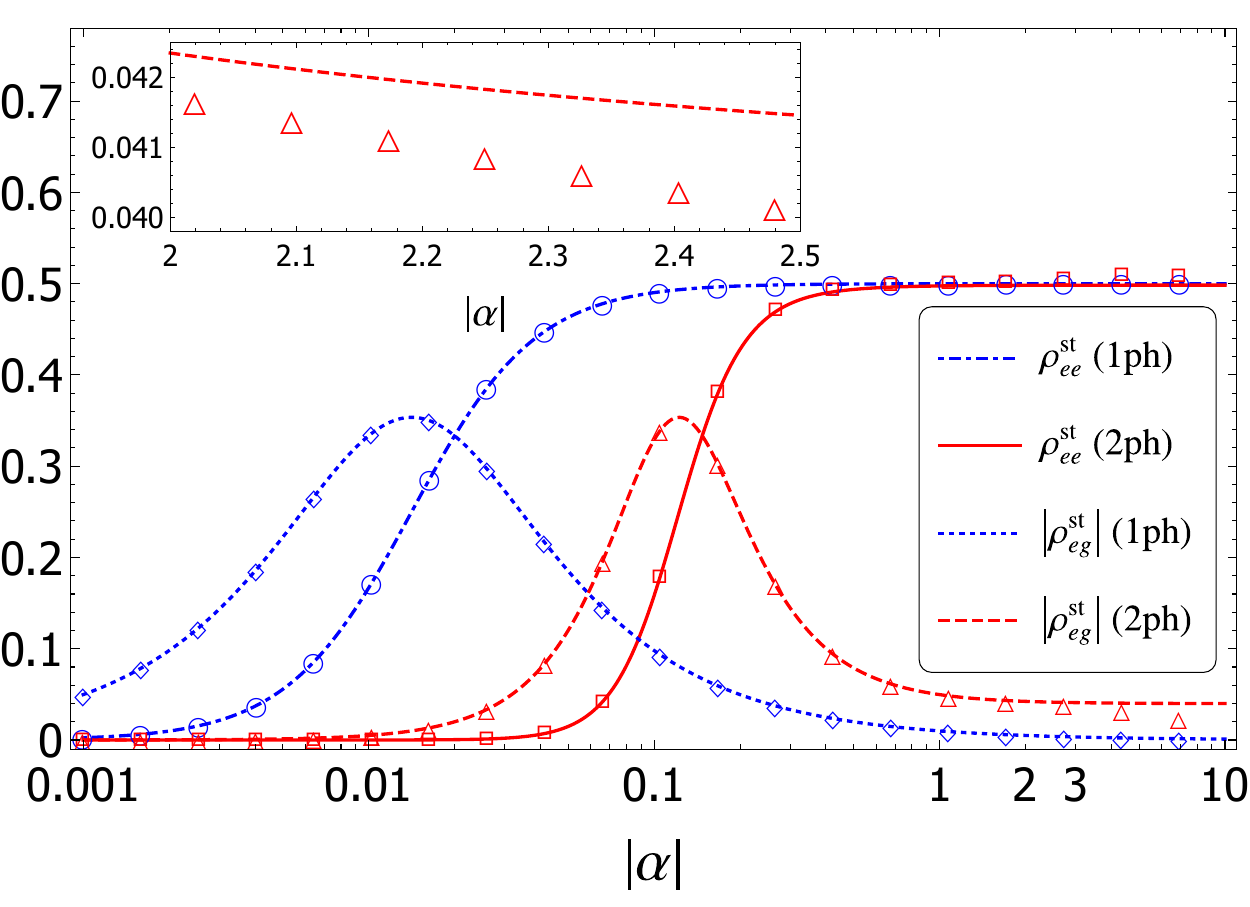}
	\caption{Steady-state excited populations and coherences of one TLS as a function of $\abs{\alpha}$ with $\gamloc=0$, $g=0.01k$, $T=0$, and $P=0$.
	The various empty markers show discrete points computed with the full model of Eq.~\eqref{eq:FullModel}.
	As predicted, the error induced by the effective model increases as $\abs{\alpha}$ increases.
	The inset shows a zoom of the 2ph steady-state coherence for $ 2 \leq \abs{\alpha} \leq 2.5 $.
	Both the full and the effective model predict a very low variation of the coherence in this range of $\abs{\alpha}$.}
	\label{fig:StableSteadyState}
\end{figure}
 
Focusing on the reachable steady states $\rho^{\textup{st}}$ in the one TLS case, Fig.~\ref{fig:StableSteadyState} shows that nondiagonal ones in the bare basis, that is, those presenting coherences, can be obtained. The analytical expression of these coherences in the general case ($T\neq 0$ and $P\neq 0$) can be found in Appendix~\ref{APPSubSec:OneTLSCase}. In particular, non negligible coherences are obtained when $g$ is sufficiently high (but inside the validity range of the adiabatic elimination). By comparing the two models, one can see that great differences arise for $\abs{\alpha} \gtrsim 1$. 
In this regime, indeed, the 2ph interaction allows  one to generate steady states in much shorter time (as one can evince from Fig.~\ref{fig:EnhancedPurcellLikeEffect}) and with higher coherences.
Moreover, the steady state does not change much for little variations of $\abs{\alpha}$ when $\abs{\alpha}$ is high enough. This is due to the fact that when $\gamloc$ is negligible, the steady state depends only on the ratio $\gamma_l/(g\abs{\alpha}^l )$, which in the 2ph case does not tend to zero but to  $16g/k$. For example, when $\gamloc=0$ and $g=0.01k$, the steady-state coherences for $2 \leq \abs{\alpha} \leq 2.5$ are very close, as shown in the inset of Fig.~\ref{fig:StableSteadyState}.
Therefore, it is possible to rapidly generate nondiagonal steady states resilient to intensity fluctuations of the coherent driving. We stress that the generation of steady-state coherence is relevant
since, in general, it is considered as a resource for quantum technologies~\cite{Streltsov2017}.
In particular, it has been recently shown that  nondiagonal steady states can find applications in quantum metrology protocols~\cite{Wang2018,Smirne2019}, which could be then enhanced by generating these states faster.

\section{Temperature resilience of collective phenomena}

Let us now consider the case of no coherent pumping, in order to focus on the emergence of correlations due to the collective dissipative terms. For $\alpha=0$, in Eq.~\eqref{eq:TheReducedMasterEquation} the unitary term disappears, $\gamma_2=\gamma_1 (1+2\nuno)$, and $
 \ndue = \nuno^2/(1+2\nuno)=1/[e^{2\hbar\omega/(k_B T)}-1]$.
This  particular setting has been used~\cite{Meiser2010II,Auffeves2011} to study the emergence of sub and superradiant steady states as a function of the incoherent pumping parameter $P$ when $T=0$.
The quantity $\Jcorr=\langle J_+ J_-\rangle - \sum_{i=1}\langle\sigma_+^{(i)}\sigma_-^{(i)}\rangle$ is used to characterize these collective phenomena. In particular, $\Jcorr >0$  indicates the occurrence of superradiance while $\Jcorr <0$ indicates that of subradiance. 

\begin{figure}
	\centering
	\includegraphics[width=0.48\textwidth]{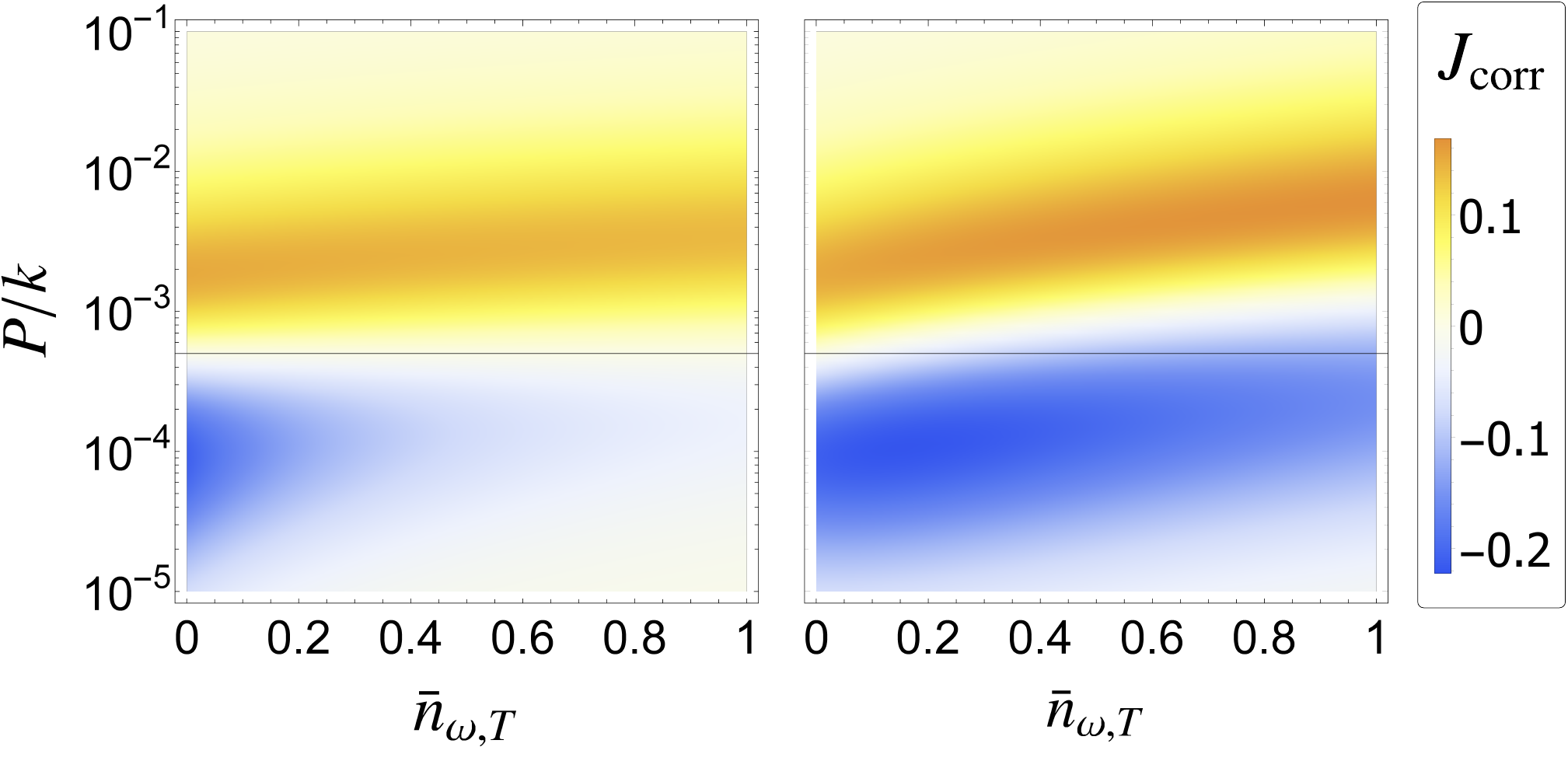}
	\caption{$\Jcorr$ of the steady state of two TLSs as a function of $T$ ($\nwt$ in the plot) and $P/k$ in the 1ph and 2ph cases for $g=0.01k$, $\gamloc = 10^{-4}k$, and $\alpha=0$.
		The horizontal lines correspond to the value  $P=P^*\equiv \gamloc + \gamma_1=5\times10^{-4}k$, where both models give exactly $\Jcorr=0$ at zero temperature.
		The 2ph model exhibits a richer dependence on temperature including stronger subradiance and superradiance at higher temperatures. Note that the extremal values that $\Jcorr$ may assume in the two-TLS case are $-1$ and $1$.}
	\label{fig:JcorrColorMap}
\end{figure}

\begin{figure}
	\centering
	\includegraphics[width=0.48\textwidth]{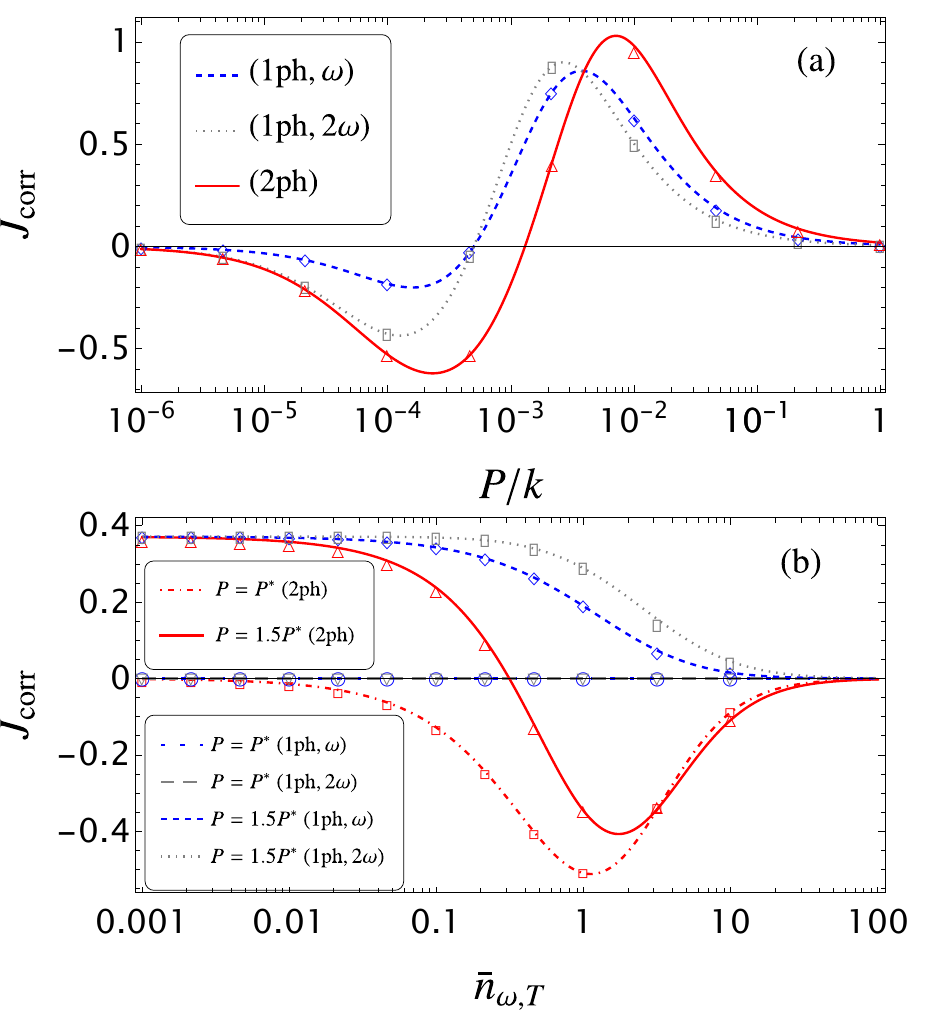}
	\caption{(a)  $\Jcorr$ of the steady state of four TLSs as a function of $P$, for $g=0.01 k$, $\gamloc = 10^{-4}k$, $T$  such that $\nwt=1$, and $\alpha=0$.
	Here, $\Jcorr$ is plotted for the 1ph (for both  $\omega$ and $2\omega$) and 2ph cases. 
	(b) Steady $\Jcorr$ of four TLSs as a function of $T$ ($\nwt$ in the plot), for $g=0.01k$, $\gamloc = 10^{-4}k$, and $\alpha=0$, for $P=P^*\equiv \gamloc + \gamma_1=5 \times 10^{-4}k $ and $P=1.5 P^*$ (see legend). The 1ph ($ \omega$ and $2 \omega$) and 2ph cases are compared.
	In both plots, $\Jcorr$ is always zero for $P=P^*$ in the 1ph case and the various empty markers indicate discrete points computed with the full model of Eq.~\eqref{eq:FullModel}, i.e., without performing the adiabatic elimination [in panel (b), because of computational difficulties only points with $\nwt$ up to 10 are considered]. Note that the extremal values that $\Jcorr$ may assume in the four-TLS case are $-2$ and $4$.}
	\label{fig:TemperatureDifferences}
\end{figure}

When $T=0$, there is no difference between the 1ph and the 2ph models because $\gamma_2 = \gamma_1$.
In contrast, the two models behave very differently  for $T\neq 0$, as shown in Fig.~\ref{fig:JcorrColorMap} where we plot the steady value of $\Jcorr$ in the two models as functions of the incoherent pumping and the baths temperature in the case of two TLSs, for $g=0.01k$ and $\gamloc = 10^{-4}k$.
A more varied dependence of the collective phenomena  on temperature in the 2ph case  is observed due to the increase of the collective dissipation rate $\gamma_2$ with the temperature. In particular, remarkable differences are observed when $P$ is close to $P^*\equiv\gamma_1 + \gamloc$, since for this value of $P$,  in the 1ph case, $\Jcorr=0$ for any $T$, while this is not the case in the 2ph case. This can be also evinced by the analytical expression we have obtained for $\Jcorr$ in the two-TLS case (see Appendix~\ref{APPSubSec:TwoTLSCase}) which shows that subradiance and superradiance are obtained when $P$ is, respectively, lower or higher than $ \gamma_l + \gamloc$. This behavior  of the sign of $\Jcorr$ has been confirmed in all the other simulations that we have done (up to six TLSs).
This means that for $P= P^*$, since $\gamma_2$ increases with temperature, subradiance is observed for any temperature different from zero in the 2ph case. One could wonder if part of these differences arises just because the TLSs in the 2ph model have frequency  $2\omega$ so that, for the same temperature, they interact with local baths by means of a lower  average excitation number. To check the extent of this effect we have also looked at the same plot using the frequency $2\omega$ for the TLSs and the HO for the 1ph case finding only a partial reduction of the differences between the two models. An example of this issue is treated for a specific example in Fig.~\ref{fig:TemperatureDifferences}.

A different behavior of collective phenomena  is still present in the case of a larger number of TLSs, as exhibited in Fig.~\ref{fig:TemperatureDifferences}(a), where the plot of $\Jcorr$ in the steady state as a function of the incoherent pumping for four TLSs at a fixed temperature ($\nwt =1$) clearly shows relevant differences in the two models, especially for the subradiance.
In particular, in the 2ph case, a higher peak of both super and subradiance can be reached, even when the frequency of the TLSs and of the HO in the 1ph case is set equal to $2 \omega$. 
A more striking different behavior of the two models can be obtained by studying the dependence of the steady value of $\Jcorr$ on $T$ for specific values of the pump, as shown in Fig.~\ref{fig:TemperatureDifferences}(b).
For $P= P^*$ no subradiance nor superradiance is visible in the 1ph case, while in the 2ph case a strong subradiance may be observed. An even more interesting case is obtained for $P > P^*$. In this case,  the system displays superradiance at $T=0$ in both models while  it follows very different paths, depending on the model,  when the temperature increases. In the 1ph model, $\Jcorr$ is always positive and tends to zero for increasing temperature whereas, in the 2ph model, there is a temperature $T'$ such that $P < \gamma_2 + \gamloc$ for $T>T'$.
Therefore, in the 2ph model, the system can go into a subradiant zone inaccessible through the 1ph interaction at fixed pumping.

\section{Conclusions}

In summary, we have studied the case of a damped HO interacting with $N$ TLSs via a two-photon coupling in the bad-cavity limit in the presence of 
finite temperature baths, a coherent pumping on the HO, and an incoherent pumping on the TLSs, comparing it to the one-photon-coupling case. We have succeeded in applying a recent adiabatic elimination technique in the two-photon model to derive a master equation governing the collective evolution of the TLSs.
This presents two fundamental differences compared to the dipolar case: an enhancement of the spontaneouslike emission rate, including a thermal contribution and a quadratic term in the coherent driving, and an increased temperature of the effective bath experienced by the TLSs.  This unexpected phenomenology makes it possible to accelerate the generation of nondiagonal one-TLS steady states and to observe a drastic change of the temperature-dependent  behavior of quantum collective phenomena, leading to a stronger resilience of these phenomena to high temperatures. We finally remark that the models here investigated can be feasibly implemented with both solid-state and atomic existing quantum technologies, as also discussed in Appendix~\ref{APPSec:CircuitModel} for the 2ph model in the solid-state context.

\begin{acknowledgements}
N.P. acknowledges the financial support of the Observatoire des Sciences de l'Univers THETA Franche-Comt\'{e} / Bourgogne for his research visit at the Universit\'{e} Paris Diderot (now Universit\'{e} de Paris). B.B. acknowledges  support by the French ``Investissements d'Avenir'' program, project ISITE-BFC (Contract No.~ANR-15-IDEX-03).
N.P. and B.B. thank Andrea Smirne for useful discussions about the results of this Letter.
\end{acknowledgements}

\appendix

\section{Circuit model\label{APPSec:CircuitModel}}

\begin{figure*}[t]
	\centering
	\includegraphics[angle=0, width=1.4\columnwidth]{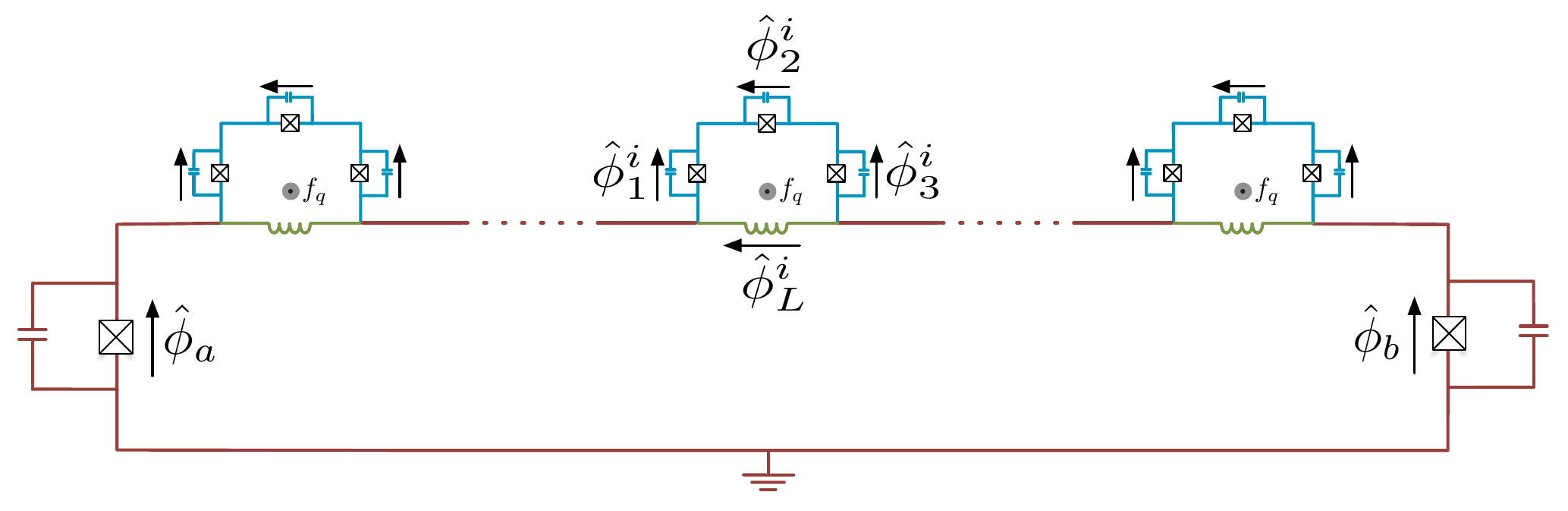}
	\caption{\label{Fig:chip} (a) Sketch of the circuit QED scheme: a  SQUID resonator (brown), coupled through linear inductive elements (green) with flux qubits (cyan). An arbitrary number of flux qubits can in principle be coupled with the SQUID resonator. For the sake of simplicity, the variables $\hat \phi_1^i$, $\hat \phi_2^i$  and $\hat \phi_3^i$, of the elements composing the flux qubits are explicitly shown only for the qubit depicted in the middle.}
\end{figure*}

In this section, we provide a detailed description of a superconducting quantum circuit scheme that can implement the two-photon (2ph) Jaynes-Cumming Hamiltonian used in Eq.~(1) of the main text for $l=2$.  Indeed, in the framework of circuit QED~\cite{blais2020circuit} it is possible to engineer artificial atoms that are nonlinearly coupled with a single-mode quantum resonator~\cite{felicetti_two-photon_2018,PhysRevA.98.053859}. We follow the derivation proposed in Ref.~\cite{PhysRevA.98.053859} for the case of one qubit, generalizing it to the multiqubit case.
Our scheme is depicted in Fig.~\ref{Fig:chip} and it consists of an arbitrary number of flux qubits~\cite{Orlando99,vanderwal00} coupled to a single superconducting quantum interference device (SQUID) resonator. For each qubit the coupling is realized via an inductance and the values of the coupling inductances are assumed to be the same for all the qubits. The SQUID is used in the linear regime so that, for the considered energy scales, it supports a quantum harmonic resonator mode. However, as shown in the following, the intrinsic nonlinearity of the device can lead to a dynamics governed by a two-photon coupling between the artificial atoms and the resonator. We will first derive an effective Lagrangian model of the circuit, and we will then quantize the corresponding Hamiltonian.

\subsection{Lagrangian}

Our starting point is the Lagrangian of the system~\cite{blais2020circuit},
\begin{equation}
	\mathcal{L}_{\rm TOT} = \mathcal{L}_{\rm SQUID} + \sum_i \left[ \mathcal{L}_{\rm FQ}^i +  \mathcal{L}_{\rm L}^i \right],
\end{equation}
where the index $i$ runs over the flux qubits and where the different contributions of the SQUID, the flux qubits (FQ), and the coupling inductances (L) are given by
\begin{equation}
	\mathcal{L}_{\rm SQUID} = \frac{C}{2}\dot \phi_a^2 +  \frac{C}{2}\dot \phi_b^2 +  E_J\left[\cos(\frac{\phi_a}{\phi_0}) + \cos(\frac{\phi_b}{\phi_0})  \right], 
\end{equation}
\begin{align}
	\mathcal{L}^i_{\rm FQ} &= \frac{\widetilde C}{2} \left[(\phidot_1^i)^2 + (\phidot_3^i)^2 \right] +  \frac{\alphaN \widetilde C}{2}(\phidot_2^i)^2 \nonumber \\
	&+ \widetilde{E}_J \left[\Cos{\frac{\phi_1^i}{\phi_0}} + \Cos{\frac{\phi_3^i}{\phi_0}} + \alphaN \Cos{\frac{\phi_2^i}{\phi_0}}  \right],  \\
	\label{SM_lagL}
	\mathcal{L}^i_{\rm L} &= - \frac{(\phi_L^i)^2}{2L}.
\end{align}
Here, $E_J$ denotes the Josephson energy and $C$ the capacitance of the Josephson junctions (JJ) $a$ and $b$ composing the SQUID, which are assumed to be identical. As for the flux qubits, $\tilde{E}_J$ and $\tilde{C}$ are the Josephson energy and capacitance of the JJs labeled by $1$ and $3$, while $\alphaN$ is a parameter, smaller than one, used to quantify the corresponding values $\alphaN \tilde{E}_J$ and  $\alphaN \tilde{C}$ for the second JJ. The coupling inductances between each flux qubit and the SQUID are also assumed to be identical with value $L$.
Finally, all the variables denoted with a $\phi$ are the generalized fluxes associated to each element of the circuit~\cite{blais2020circuit}, and we have defined the reduced magnetic flux quantum as $\phi_0 = \Phi_0/(2\pi) = \hbar/(2e)$, where $e$ is the electron charge. 

We now define symmetric and anti-symmetric SQUID variables as
$\phi_+ = \frac{\phi_a + \phi_b}{2}$ and $\phi_- = \frac{\phi_a - \phi_b}{2}$ and 
apply the flux-quantization rule~\cite{tinkham2004introduction,blais2020circuit} to the SQUID loop, obtaining $\phi_a - \phi_b = \sum_i \phi^i_L + \phi^{\rm ext}_s $, where $\phi^{\rm ext}_s$ is the external magnetic flux flowing through the SQUID loop.
We can then rewrite the anti-symmetric variable in terms of the  phases $\phi^i_L$ of the coupling inductances obtaining $\phi_- = \sum_i \frac{\phi^i_L}{2} + \frac{\phi^{\rm ext}_s}{2}$. We consider a constant external flux implying $\dot \phi^{\rm ext}_s=0$. Straightforward  calculations allow us to rewrite the SQUID Lagrangian as,
\begin{align}
	\label{SM_tobelinearized1}
	\mathcal{L}_{\rm SQUID} =& C \phidot_+^2 + \frac{C}{4} (\phidot_L^{\rm tot})^2 \nonumber \\
	&+ 2 E_J\Cos{\frac{\phi^{\rm tot}_L + \phi^{\rm ext}_s}{2\phi_0}}\Cos{\frac{\phi_+}{\phi_0}},
\end{align}
where we have defined $\phi^{\rm tot}_L = \sum_i \phi^i_L$.
Now, we turn our focus on the circuit elements composing the flux qubits, and we define $\phi_p^i = \frac{\phi_1^i + \phi_3^i}{2}$  and $\phi_m^i = \frac{\phi_1^i - \phi_3^i}{2}$. We denote with $\phi^{\rm ext}_q$ the external flux on each qubit, which is taken to be the same for all flux qubits, and which is defined in the opposite direction with respect to $ \phi^{\rm ext}_s$. The  flux-quantization rule for the qubit loop, $\phi_1^i - \phi_2^i - \phi_3^i = -  \phi_L^i - \phi^{\rm ext}_q $, allows us to eliminate the phase variable of the second junction of each qubit: $\phi_2^i = 2\phi_m^i + \phi^i_L + \phi^{\rm ext}_q$. We take also the external flux biasing the flux qubit to be constant implying $\dot \phi^{\rm ext}_q = 0$, so we can write
\begin{align}
	\label{SM_tobelinearized2}
	&\mathcal{L}^i_{\rm FQ} =  \widetilde{C} (\phidot_p^i)^2 + \widetilde{C} (\phidot_m^i)^2 + \frac{\alphaN\widetilde C}{2}\left( 2\phidot^i_m + \phidot^i_L \right)^2 + \nonumber \\  
	&\widetilde{E}_J \!\left[\! 2\Cos{\frac{\phi_p^i}{\phi_0} }\!\Cos{\frac{\phi_m^i}{\phi_0} }\! +\! \alphaN \Cos{\frac{2\phi_m^i + \phi_L^i +  \phi^{\rm ext}_q}{\phi_0}}  \!\right]\!.  
\end{align}

We now take a perturbative approach based on the assumption that the coupling inductances are chosen to be so small that the following approximations are justified. In particular, we perform two main approximations: (1) we linearize the Lagrangian with respect to the coupling-inductance phase variables divided by $\phi_0$, that is we expand to first order in  $\phi^i_L /\phi_0$, and  then (2) we perform an adiabatic elimination of the corresponding degrees of freedom. Notice that this adiabatic elimination concerns only the detail of the circuit design and it is not related with the adiabatic elimination performed on the quantum model considered in the main text.
To simplify the notation, let us define the gauge-invariant phase variables $\varphi_j= \phi_j/\phi_0$, where $\phi_j$ denotes generically the flux variable of any circuit element. We also define the frustration parameters $f_s = \phi^{\rm ext}_s/\phi_0$ and $f_q = \phi^{\rm ext}_q/\phi_0$.

(1) First, we assume that the flux variables $\phi^i_L$ are small with respect to the reduced magnetic flux quantum, so that Eq.\eqref{SM_tobelinearized1} and Eq.\eqref{SM_tobelinearized2} can be linearized with respect to the variables $\varphi_L^i$. Thus, from  Eq.~\eqref{SM_tobelinearized1} we obtain
\begin{multline}
	\label{SM_lagSQUID}
	\mathcal{L}_{\rm SQUID} = C \phidot_+^2 + \frac{C}{4} \prt{\phidot_L^{\rm tot}}^2   \\
	+ 2 E_J \left[\Cos{\frac{f_s}{2}} - \Sin{\frac{f_s}{2}}\frac{\varphi_L^{\rm tot}}{2} \right] \Cos{\varphi_+}. 
\end{multline}
Then, by linearizing Eq.\eqref{SM_tobelinearized2} with respect to each $\varphi_L^i$ we obtain,
\begin{align}
	\label{SM_lagFQ}
	\mathcal{L}^i_{\rm FQ} =&  \mathcal{L}^i_{\rm qubit}  + 
	\frac{\alphaN\widetilde C}{2}\left[\left(\phidot_L^i\right)^2 + 4 \phidot_L^i\phidot_m^i \right] \nonumber \\
	&+\alphaN \widetilde{E}_J \Sin{2\varphi_m^i+f_q}\varphi_L^i,
\end{align}
where $ \mathcal{L}^i_{\rm qubit}$ denotes the standard Lagrangian of the $i$-th flux qubit~\cite{vanderwal00,Orlando99},
\begin{multline}
	\mathcal{L}^i_{\rm qubit} = \widetilde{C} \prt{\phidot_p^i}^2 + \left(1+2\alphaN \right)\widetilde{C}  \prt{\phidot_m^i}^2   \\ 
	+ \widetilde{E}_J \left[ 2\Cos{\varphi_p^i }\Cos{\varphi_m^i } + \alphaN \Cos{2\varphi_m^i + f_q}  \right].
\end{multline}

(2) Now, we perform the adiabatic elimination on the degrees of freedom of the coupling inductances. 
These inductances appear in the following terms of the total Lagrangian [see Eqs.~\eqref{SM_lagL}, \eqref{SM_lagSQUID}, and \eqref{SM_lagFQ}]:
\begin{align}
	&\frac{1}{2}\!\prtq{\!\prt{\frac{C+2\alphaN\widetilde C}{2}}\! \sum_i  \prt{\phidot^i_L}^2 \!- \frac{1}{L}\sum_i \prt{\phi_L^i}^2\!
		+ \frac{C}{2} \sum_{i\neq j} \phidot^i_L \phidot^j_L}\nonumber\\
	&- E_J \Sin{\frac{f_s}{2}} \sum_i \phi_L^i \Cos{\varphi_+}\nonumber\\
	&+ 2 \alphaN \widetilde{C} \sum_i \phidot_L^i \phidot_m^i
	+ \alphaN \widetilde{E}_J \sum_i \varphi_L^i \Sin{2 \varphi_m^i + f_q}.
\end{align}
The first line corresponds to $N$ interacting harmonic oscillators. Given the high symmetry of the coefficients, it is easy to find the normal frequencies~\cite{BookLandau1976Mechanics}, which are given by
\begin{equation}
	\label{eq:NormalFrequencies}
	\omega_{L}^- = \sqrt{\frac{1}{L\prt{\alphaN \widetilde{C} + NC/2}}},
	\qquad
	\omega_{L}^+ = \sqrt{\frac{1}{L \alphaN \widetilde{C}}},
\end{equation}
where the frequency $\omega_{L}^+$ has degeneracy $N-1$.
Here, $L$ is assumed to be so small that the frequency $\omega_{L}^-$ (which is clearly smaller than $\omega_{L}^+$) is much larger than the relevant characteristic frequencies of the other elements of the circuit.
Moreover, the interaction of the inductances with the other elements of the circuit is such that, if $L$ is small enough, we can adiabatically eliminate the variables corresponding to the inductances by setting  $\phidot^i_L = 0$.

To simplify the notation let us define the following parameters:
\begin{equation}
	S= E_J\Sin{\frac{f_s}{2}},\quad
	\Omega_m^i = -\alphaN \widetilde E_J\Sin{2\varphi_m^i + f_q}.
\end{equation}
Imposing $\phidot^i_L = 0$ in the Euler-Lagrange equation
\begin{equation}
	\pdv{\phi_L^{i}} \mathcal{L}_{\rm TOT} - \dv{t} \pdv{\phidot_L^i}   \mathcal{L}_{\rm TOT} = 0,
\end{equation}
we obtain the dependence of $\phi^i_L$ on the remaining dynamic variables,
\begin{equation}
	\label{fiel}
	\varphi_L^i = \frac{\phi^i_L}{\phi_0} = -\frac{1}{2 E_L}\prtq{S \cos(\varphi_+) + \Omega_m^i},
\end{equation}
where $E_L = \phi_0^2/(2L)$. Therefore,
\begin{equation}
	\label{fieltot}
	\varphi_L^{\rm tot} = \frac{1}{\phi_0}\sum_i \phi^i_L = -\frac{1}{2 E_L}\prtq{S N \cos(\varphi_+) + \sum_i \Omega_m^i}, 
\end{equation}
where $N$ is the total number of flux qubits.

Before replacing the variables, in order to simplify the expressions, we also define
\begin{equation}
	\label{SM_pardef}
	K = 2E_J \Cos{\frac{f_s}{2}}.
\end{equation}
Within the above approximations, the total Lagrangian is then obtained adding Eqs.~\eqref{SM_lagSQUID}, \eqref{SM_lagFQ} and  \eqref{SM_lagL}, and it can be written as,
\begin{align}
	\label{SM_LagphiL}
	\mathcal{L}_{\rm TOT} =& C\phidot_+^2 +  \left[ K-  S \varphi_L^{\rm tot} \right] \Cos{\varphi_+}   \nonumber \\
	&+\sum_i \left[ \mathcal{L}^i_{\rm qubit} - \Omega_m^i \varphi^i_L - E_L(\varphi_L^i)^2 \right].
\end{align}

Finally,  by replacing $\varphi^i_L$ and $\varphi_L^{\rm tot}$ by means of, respectively, Eqs.~\eqref{fiel} and \eqref{fieltot} we obtain,
\begin{multline}
	\label{A_lagr}
	\mathcal{L}_{\rm TOT} =  C\phidot^2_+ + K \Cos{\varphi_+} + \frac{N S^2}{4E_L}\cos[2](\varphi_+)   \\ 
	+ \frac{S \cos(\varphi_+)}{2 E_L} \sum_i \Omega_m^i + \frac{1}{4 E_L}\sum_i \prt{\Omega_m^i}^2 + \sum_i \mathcal{L}_{\rm qubit}^i.
\end{multline}

\subsection{Hamiltonian}
The system Hamiltonian $\mathcal{H}_{\rm TOT}$ can be derived 
implementing the Legendre transformation, i.e.,
\begin{equation}
	H_{\rm TOT} = \varphidot p_+ + \sum_j \varphidot_p^j p_p^j + \sum_j \varphidot_m^j p_m^j- \mathcal{L}_{\rm TOT},
\end{equation}
where we use the standard definition of conjugate variables
\begin{align}
	p_+ &= \partial \mathcal{L}_{\rm TOT}/\partial \varphidot_+ = 2 C \phi_0^2 \varphidot_+, \nonumber \\
	p_p^j &= \partial \mathcal{L}_{\rm TOT}/\partial \varphidot_p^j 
	= 2 \tilde{C} \phi_0^2 \varphidot_p^j, \nonumber \\
	p_m^j &= \partial \mathcal{L}_{\rm TOT}/\partial \varphidot_m^j
	= 2 (1 + 2 \alphaN) \tilde{C} \phi_0^2 \varphidot_p^j.
\end{align}

We replace now the classical variables with quantum operators and we start using the hat formalism to avoid confusion. The total Hamiltonian can be written as
\begin{equation}
	\label{SM_fullHam}
	\hat{H}_{\rm TOT} = \hat{H}_{\rm SQUID} + \sum_i \hat{H}_{\rm FQ}^i + \hat{H}_I.
\end{equation}
The SQUID Hamiltonian is given by
\begin{equation}
	\label{SM_SquidHam}
	\hat{H}_{\rm SQUID} = \frac{\hat p_+^2}{4C\phi_0^2} -  K \Cos{\hat  \varphi_+} - \frac{N S^2}{4E_L}\cos[2](\hat  \varphi_+).
\end{equation}
The Hamiltonian $\hat{H}_{\rm FQ}^{i}$ is given by the standard flux-qubit Hamiltonian $\hat{H}_{\rm FQ}^{i, {\rm st}}$, plus a correction proportional to the small parameter $L$ [since $E_L = \phi_0^2 /(2L)$]:
\begin{equation}
	\label{SM_fluxqubit_first}
	\hat{H}_{\rm FQ}^i =\hat{H}_{\rm FQ}^{i, {\rm st}} - \frac{(\hat  \Omega_m^i)^2}{4E_L},
\end{equation}
where
\begin{align}
	\label{StandardFluxQubit}
	&\hat{H}_{\rm FQ}^{i, {\rm st}}
	= \frac{(\hat p_p^i)^2}{4 \widetilde C \phi_0^2 } +  \frac{(\hat p_m^i)^2}{4\widetilde C \phi_0^2  (1+2\alphaN)}  \nonumber \\ 
	&- \widetilde{E}_J \left[ 2\Cos{\hat \varphi_p^i }\Cos{\hat \varphi_m^i } +\alphaN \Cos{2\hat \varphi_m^i + f_q}  \right].
\end{align}
It is well known that the standard flux-qubit Hamiltonian has a strongly nonlinear eigenspectrum~\cite{Orlando99,vanderwal00}. As a result, for the relevant energy scale, the system dynamics is constrained in the lowest two-level subspace~\cite{Orlando99,vanderwal00}. In the next subsection we discuss the role of the additional term $-(\hat  \Omega_m^i)^2/(4E_L)$, and we show that it does not induce state leakage outside the low-energy subspace so that each flux qubit can indeed be considered as a two-level system (TLS). In the following, the two states of the low-energy subspace for the $i$-th TLS are denoted as $\ket{0}_i$ and $\ket{1}_i$ and the Hamiltonian of each TLS is $\hbar\omega_q\hat{\sigma}_z^i/2$, where $\omega_q$ is the common transition frequency of the TLSs and $\hat{\sigma}_z^i$ is the usual $z$-Pauli operator.

The last term in Eq.~\eqref{SM_fullHam} corresponds to the nondipolar coupling Hamiltonian
\begin{equation}
	\label{SM_nondipcoup_1}
	\hat{H}_I = -\frac{S}{2E_L} \Cos{\hat \varphi_+} \sum_i \hat \Omega^i_m.
\end{equation}
We show in the following that, in a broad regime of parameters, such nondipolar coupling can be reduced to a two-photon interaction plus an additional correction to the flux-qubit Hamiltonian.

\subsection{Effective model}
We now assume that the phase of the SQUID junctions is small compared to the reduced magnetic flux quantum: $\varphi_+ = \phi_+/\phi_0\ll 1$. This is a standard approximation~\cite{blais2020circuit} which is valid when the SQUID operates in the linear regime, that is in the limit of large Josephson energy for the two Josephson junctions, $a$ and $b$, of the SQUID.
Expanding up to second order the cosines and discarding constant terms in Eq.~\eqref{SM_SquidHam} we obtain,
\begin{equation}
	\label{SM_SQUIDexp}
	\hat{H}_{\rm SQUID} = \frac{\hat p_+^2}{4\phi_0^2C} +  \left(K + \frac{N S^2}{2E_L} \right)  \frac{\hat  \varphi_+^2}{2}.
\end{equation}
Similarly, we obtain from Eq.~\eqref{SM_nondipcoup_1}
\begin{equation}
	\label{SM_nondipcoup_2}
	\hat{H}_I = - \frac{S}{2E_L} \sum_i \hat \Omega^i_m + \frac{S}{2E_L} \frac{\hat  \varphi_+^2}{2} \sum_i \hat \Omega_m^i,
\end{equation}
where the first term is a free energy term of the qubit, while the second term is the origin of the nondipolar coupling.

We now introduce the standard ladder operators $\hat a$ and $\hat a^\dagger$ of the quantum harmonic oscillator corresponding to the SQUID Hamiltonian of Eq.~\eqref{SM_SQUIDexp} in
\begin{equation}
	\label{SM_def_a_adag}
	\hat \varphi_+ = \sqrt{\frac{\hbar \omega_c L_{\rm eff}}{2\phi_0^2}}\left( \hat a^\dagger+ \hat a\right),\quad 
	\hat p_+ = i \sqrt{\frac{\hbar \phi_0^2}{2\omega_c L_{\rm eff}}}\left( \hat a^\dagger - \hat a\right),
\end{equation}
where we have defined
\begin{align}
	L_\eff =& \frac{\phi_0^2}{\left( K + \frac{N S^2}{2E_L} \right)}, \nonumber \\
	\omega_c =& \sqrt{\frac{1}{  2 C  L_\eff}} = \frac{1}{\hbar}\sqrt{4E_C\left(K + \frac{N S^2}{2E_L} \right)},
\end{align}
where we have introduced the charging energy $E_C = e^2/(2C)$ [we recall that $\phi_0 = \hbar /(2 e)$].
Equation~\eqref{SM_SQUIDexp} can be then rewritten as
\begin{equation}
	\label{SM_rabi1}
	\hat{H}_{\rm SQUID} = \hbar \omega_c \hat a^\dagger \hat a, 
\end{equation}
where we have used $[\hat a,\hat a^\dagger]=1$ and we have again disregarded constant terms.

Now, we redefine the total free Hamiltonian of a flux qubit as the sum of Eq.~\eqref{SM_fluxqubit_first} and of the $i$-th element of the first term in Eq.~\eqref{SM_nondipcoup_2},
\begin{equation}
	\label{SM_rabi2}
	\hat{H}_{\rm FQ}^i = \hat{H}_{\rm FQ}^{i, {\rm st}}
	- \left[  \frac{1}{4E_L}\prt{\hat \Omega_m^i}^2  +  \frac{S}{2E_L}\hat \Omega^i_m  \right],
\end{equation}
which corresponds to the standard Hamiltonian of a flux qubit plus two corrections. In the first two energy levels subspace we can write~\cite{Orlando99,vanderwal00} 
$\hat \Omega^i_m =  -\alphaN \widetilde E_J \bra{0}_i \Sin{2\hat \varphi_m^i + f_q} \ket{1}_i \hat \sigma_x^i = - \alphaN \widetilde E_J T(f_q) \hat \sigma_x^i $, where $\hat{\sigma}_x^i$ is the usual $x$-Pauli operator and the transition amplitude $T(f_q)$ is the same for all the qubits because we have set a homogeneous $f_q$ and the operator $\hat \varphi_m^i$ has the same form for all qubits.
Notice that the first correction, proportional to $(\hat\Omega_m^i)^2 $, corresponds to a constant energy offset, since $(\hat\Omega_m^i)^2 \propto \Id$ in the two-level subspace, being $\Id$ the identity operator in a two-dimensional Hilbert space, and can be then disregarded. The second one, proportional to $\hat\Omega_m^i$, can be compensated by a small adjustment  of the frustration parameter $f_q$ leading to a renormalization of the qubit-cavity coupling~\cite{vanderwal00,Orlando99}. Therefore, these additional terms do not modify the behavior of the flux qubit, and so the latter can be faithfully modelized as a TLS. In this limit, the Hamiltonian of a flux qubit is then redefined as
\begin{equation}
	\label{SM_TLS}
	\hat{H}_{\rm FQ}^i = 
	\frac{\hbar \omega_q}{2} \hat{\sigma}_z^i.
\end{equation}

Finally, we redefine the interaction Hamiltonian as given by only the second term in Eq.~\eqref{SM_nondipcoup_2}, which corresponds to the nondipolar interaction Hamiltonian between the TLSs and the resonator mode,
\begin{equation}
	\label{SM_rabi3}
	\hat{H}_I = \hbar g_2 \left(\hat a^\dagger + \hat a \right)^2  \sum_i \hat{\sigma}_x^i,
\end{equation}
where we have defined the two-photon coupling strength $g_2$ as
\begin{equation}\label{g2}
	g_2 =  \frac{S}{4 \hbar E_L}\sqrt{\frac{E_C}{ \left(K + \frac{N S^2}{2E_L} \right)}} \alphaN \widetilde E_J T(f_q).
\end{equation}

Now, we consider the resonant case by setting $\omega \equiv \omega_c=\omega_q/2$  and we redefine for this case the total system Hamiltonian $\hat{H}_{\rm TOT}$ up to second order in $\hat{\varphi}_+$ as the sum of Eqs.~\eqref{SM_rabi1}, \eqref{SM_TLS}, and \eqref{SM_rabi3}, obtaining
\begin{equation}
	\label{SM_TotalRabiHamiltonian}
	\hat{H}_{\rm TOT} = \hbar \omega \hat{a}^\dagger \hat{a} + \hbar \omega \hat{J}_z + \hbar g \prt{\hat{a} + \hat{a}^\dagger}^2\prt{\hat{J}_+ + \hat{J}_-},
\end{equation}
where $g \equiv g_2$, $\hat{J}_z = \sum_i \hat{\sigma}_z^{(i)}$, and $\hat{J}_\pm = \sum_i \hat{\sigma}_\pm^{(i)}$, as in the main text, being $\hat{\sigma}_+^{(i)}$ and $\hat{\sigma}_-^{(i)}$  the usual raising and lowering operators for a TLS.
Notice that, as shown in Ref.~\cite{PhysRevA.98.053859}, the fourth-order corrections here neglected have a negligible impact on the system dynamics and spectral features. 

Going to the interaction picture through the unitary operator $\hat{U}_{0} = \exp{-i \hat{H}_0 t/\hbar}$, where $\hat{H}_0 = \hbar \omega \hat{a}^\dagger \hat{a} + \hbar \omega \hat{J}_z$, the interaction Hamiltonian becomes $\hat{H}_{I, \mathrm{int}}=\hat{U}_{0}^\dagger \hat{H}_I \hat{U}_{0}$, with
\begin{multline}
	\label{SM_InteractionPictureInteraction}
	\hat{H}_{I, \mathrm{int}}/(\hbar g) = \prt{\hat{a}^\dagger}^2 \hat{J}_- + \hat{a}^2 \hat{J}_+\\
	+ e^{2 i \omega t}\prtq{\prt{2 \hat{a}^\dagger \hat{a} + 1}\hat{J}_+} + e^{- 2 i \omega t}\prtq{\prt{2 \hat{a}^\dagger \hat{a} + 1}\hat{J}_-} \\
	+ e^{4 i \omega t}\prt{\hat{a}^\dagger}^2 \hat{J}_+ + e^{- 4 i \omega t} \hat{a}^2 \hat{J}_-.
\end{multline}
For the regime of the parameters we explore in the main text (where $g \ll \omega$), the ubiquitous rotating-wave approximation (RWA) can be applied to Eq.~\eqref{SM_TotalRabiHamiltonian} and \eqref{SM_InteractionPictureInteraction}, neglecting all the terms oscillating in the interaction picture. We can thus redefine the total Hamiltonian as the two-photon Jaynes-Cumming Hamiltonian used in Eq.~(1) of the main text for $l=2$ (we remark that in the main text and in the following sections of the supplemental material the operators are not anymore marked by the \enquote{hat}):
\begin{equation}
	\label{SM_TotalJC}
	\hat{H}_{\rm TOT} = \hbar \omega \hat{a}^\dagger \hat{a} + \hbar \omega \hat{J}_z + \hbar g \prtq{\hat{a}^2  \hat{J}_+ + \prt{\hat{a}^\dagger}^2\hat{J}_-}.
\end{equation}
An analysis of the RWA in an analogous context can be found in~\cite{felicetti_spectral_2015}.

{\renewcommand{\arraystretch}{2.2}%
	\begin{table}[]
		\resizebox{0.48\textwidth}{!}{%
			\begin{tabular}{|c|c|c|c|}
				\hline
				\multicolumn{4}{|c|}{\Large Qubit parameters} \\ \hline 
				$\tilde{E}_J/h= 200$ GHz &
				$\tilde{E}_C/h = \tilde{E}_J/(80 h) = 2.5$ GHz &
				$\alphaN = 0.8$ &
				$f_q/(2 \pi)= 0.485$ \\ \hline
				\multicolumn{4}{|c|}{\Large Resonator and inductances parameters} \\ \hline
				$E_J/h = 13$ GHz &
				$E_C/h = 1$ GHz &
				$E_L/h = 2.5 \times 10^4$ GHz &
				$f_s/(2 \pi) = 0.1$ \\ \hline
				\multicolumn{4}{|c|}{ \Large Resulting parameters}                 \\ \hline
				$\omega_c/(2\pi) \approx 10 $ GHz &
				$\omega_q/(2 \pi) \approx 20 $ GHz &
				$g_2 \approx 10^{-4} \omega_c$ &
				$\omega_L^-/(2 \pi) \approx 415 $ GHz \\ \hline
			\end{tabular}%
		}
		\caption{This table reports an example of circuit parameters able to implement the model of the main text in the case of four TLSs, i.e., for $N=4$ [we have defined $\tilde{E}_C=e^2/(2\tilde{C})$]. The resulting parameters are also reported. Notice the use of the non-reduced Planck's constant $h$ for quantifying energies.
			The parameters chosen for the flux qubit have been taken from an example of Ref.~\cite{Orlando99} on the basis of the standard flux qubit  Hamiltonian $\hat H_{\rm FQ}^{i, {\rm st}}$. Notice that, because of the two additional terms in Eq.~\eqref{SM_rabi2} with respect to $\hat H_{\rm FQ}^{i, {\rm st}}$, the parameter values should be slightly calibrated in a real experimental realization in order to obtain the resonance $\omega_q = 2 \omega_c$. Regarding the number $T(f_q)$ which appears in the formula for $g_2$ of Eq.~\eqref{g2}, we have used $T(f_q)=0.8$. This value has been calculated for similar circuit parameters in the code used for Ref.~\cite{PhysRevA.98.053859}. We remark that, to implement our model, only the order of magnitude of $g_2$ is important, not its exact value.}
		\label{Table:CircuitQuantities}
\end{table}}

To conclude this section, let us provide an example of a set of physical parameters that matches the required regime. We consider a system composed of 4 TLSs, as in the case of Fig.~4 of the main text. Notice that all model parameters can be analytically derived, except for the flux-qubit frequency which has been taken by an example of Ref.~\cite{Orlando99}, where the same values for the flux-qubit circuit parameters have been used. The values of the various parameters are summarized in Table~\ref{Table:CircuitQuantities}, which has been constructed as follows. First, we have taken the flux-qubit circuit parameters from Ref.~\cite{Orlando99}, where the resulting frequency of the flux qubit is also given: $\omega_q/(2\pi) \approx 20$ GHz. Then, we have searched for a configuration of the resonator and inductances parameters ($E_J$, $E_C$, $E_L$, and $f_s$) leading to $\omega_c/(2\pi) \approx 10$ GHz for the harmonic oscillator, well within the working range of circuit QED devices~\cite{blais2020circuit}. For the number $T(f_q)$ appearing in the expression for $g_2$ of Eq.~\eqref{g2}, we have used the value $T(f_q)=0.8$, which has been calculated for similar circuit parameters in the code used for Ref.~\cite{PhysRevA.98.053859}.
Regarding the inductances, from Eq.~\eqref{fiel} one can estimate by excess the expectation value of the operator associated to the variable $\varphi^i_L$ finding in modulus $\approx 0.0026$, which is consistent with the linearization procedure we have performed with respect to the variables $\varphi^i_L$. Morevoer, this choice of the circuit parameters leads to a value of the lowest inductances characteristic frequency $\omega_L^-$ well above the relevant energy scale, being $\omega_L^-$ more than 20 times larger than the flux qubits frequency~\footnote{For convenience of the reader, we report in this note the formulas for the inductances normal frequencies in terms of the energetic circuit parameters:
	\begin{equation*}
		\omega_{L}^- = \frac{4 \sqrt{2}}{\hbar} \sqrt{\frac{E_C \tilde{E}_C E_L}{N \tilde{E}_C + 2 \alphaN E_C}},
		\quad \omega_{L}^+ =\frac{4}{\hbar} \sqrt{\frac{\tilde{E}_C E_L}{\alphaN}}.
	\end{equation*}
	These formulas are equivalent to those of Eq.~\eqref{eq:NormalFrequencies}.}.
It follows that the adiabatic elimination we have made concerning the variables $\phi^i_L$, $\dot{\phi}^i_L=0$, is well justified. 
Finally, we have $g_2 \approx 10^{-4}\omega_c$, that is $g_2 \approx 0.01k$ (this is the value considered in all the figures of the main text) if we take $k = 0.01 \omega_c$, where $k$ is the dissipation rate of the harmonic oscillator due to the interaction with its environment.

Concerning the flux qubits, the parameters are taken from Ref.~\cite{Orlando99} on the basis of the standard flux qubit  Hamiltonian $\hat{H}_{\rm FQ}^{i, {\rm st}}$, i.e., the Hamiltonian of Eq.~\eqref{SM_rabi2} without the two additional terms. It follows that the parameter values should be slightly calibrated in a real experimental realization in order to obtain the resonance $\omega_q = 2 \omega_c$. However, the engineering of flux qubit is a well-developed area of research and their effective frequency can be finely tuned in various ways, also adding additional elements to the circuit~\cite{blais2020circuit}. Even if the values of Table~\ref{Table:CircuitQuantities} do not exactly match those that would be used for our circuit, they are then a good indicator of the experimental values that would be in fact necessary.

We stress out that the values used in Table~\ref{Table:CircuitQuantities} are commonly implemented in nowadays experiments~\cite{blais2020circuit} and that the proposed circuit design does not require any further improvement over standard circuit-QED technology.

\section{Adiabatic elimination technique\label{APPSec:AdiabaticEliminationGeneral}}

Here, we briefly resume the recently developed adiabatic elimination technique~\cite{Azouit2017,ThesisAzouit2017} that we have exploited in our analysis.
To apply this technique we have to divide the system under study into two subsystems, one governed by a fast dynamics and the other one by a slow one.
We call \enquote{system $A$} the fast system and \enquote{system $B$} the slow one.
Moreover, the two subsystems have to interact weakly.
As final requirement, system $A$ has to converge to a unique steady state, which we call $\cst$, when it is influenced uniquely by its own Lindbladian.
For a far more detailed discussion see Ref.~\cite{ThesisAzouit2017}.

Following the prescription of Ref.~\cite{ThesisAzouit2017}, the Lindbladian describing the evolution of the density matrix of the global system, $\rho_G$, is given by
\begin{equation}
	\label{eq: adiabatic elimination starting point}
	\dot{\rho}_G=\Lb_A (\rho_G) + \epsilon \Lb_B (\rho_G) - 
	\frac{i}{\hbar} \epsilon \comm{\intH}{\rho_G},
\end{equation}
where $\epsilon$ is the quantity which will play the role of perturbative parameter in the expansion of system $B$ dynamics and each Lindbladian is of the form
\begin{equation}
	\Lb_{r} (\bullet) =  - \frac{i}{\hbar} \comm{H_{r}}{\bullet} + \sum_n \mathcal{D}_{X_{r}^{(n)}}(\bullet),
\end{equation}
where $r=A$, $B$, and
\begin{equation}
	\mathcal{D}_{X_{r}^{(n)}} (\bullet)=  X_{r}^{(n)}\bullet X_{r}^{\dagger (n)} - \frac{1}{2} \acomm{X_{r}^{\dagger (n)} X_{r}^{(n)}}{\bullet}.
\end{equation}
The interaction Hamiltonian can be written in the general form $\intH = \hbar c \sum_{k=1}^M A_k \otimes B^\dagger_k$, where $A_k$ and $B_k$ are not necessarily hermitian and $c$ is a constant with the dimension of a frequency.

The goal of the adiabatic elimination procedure is to find the superoperator describing the dynamics of the reduced density matrix of system $B$, $\rho_B=\Tr_A{\{\rho_G\}}$, as
\begin{equation}
	\dot{\rho}_B=   \Lb_S (\rho_B)= \sum_{m \geq 0} \epsilon^m \Lb_{S,m} (\rho_B),
\end{equation}
and to be able to return back to the global dynamics through
\begin{equation}
	\rho_G = \K (\rho_B) = \sum_{m \geq 0} \epsilon^m \K_m (\rho_B),
\end{equation}
where, at any order in $\epsilon$, $\Lb_S$ is a Lindbladian and $\K$ is a Kraus map.
In our case, we want to obtain a second order equation for the dynamics of system $B$ since, at first order, the adiabatic elimination does not give rise to dissipative terms.

Due to the peculiarities of the method employed, the zero-order terms can be chosen with a certain arbitrariness. Following Ref.~\cite{ThesisAzouit2017}, the simplest choice is $\Lb_{S,0} (\rho_B) = 0$ and $\K_0(\rho_B) = \cst \otimes \rho_B$.
Then, the first order reduced dynamics is given by \cite{ThesisAzouit2017}
\begin{equation}\label{eq: first order reduced dynamics}
	\dot{\rho}_B = \epsilon \Lb_{S,1} (\rho_B),
\end{equation}
where
\begin{equation}\label{eq: first order reduced dynamics2}
	\Lb_{S,1} (\rho_B) = - i c \sum_{k=1}^M \comm{\Tr_A \prtg{A_k \cst} B_k^\dagger}{\rho_B } + \Lb_B (\rho_B).
\end{equation}
The superoperator $\K_1$ can be obtained as follows~\cite{ThesisAzouit2017}:
\begin{equation}
	\K_1 (\rho_B) = -i c \sum_{k=1}^M F_k \prt{\cst} \otimes \prt{B_{k}^\dagger \rho_B}+ \hc\,,
\end{equation}
where $F_k \prt{\cst}= \tau \prtq{\J_A \prtT{A_k \cst} - \Tr\prtgT{A_k \cst}\cst}$, h.c.  indicates the hermitian conjugate, and $\J_A$ and $\tau$ are defined in the following.
In general, the $\J_A$ superoperator has the form
\begin{equation}
	\J_A (Z) = \frac{1}{\tau}\int_0^{\infty} e^{t\Lb_A} \prtq{Z-\R (Z)}\dd{t} + \R(Z),
\end{equation}    
where $ \tau > 0$ such that $ -\Lb_A \prt{\tau \J_A (Z)} = Z - \R(Z)$ and $ R(Z) = \lim_{s\rightarrow +\infty} e^{s \Lb_A} (Z)
= \Tr_A \{Z\} \cst$.
Notice that even if it could seem that the results of the procedure depend on the choice of the parameter $\tau$, for the case we are going to examine the value of this parameter will be irrelevant.

In order to find the second order dynamics of system $B$, it is useful to define  two matrices whose elements are given by
\begin{align}
	\label{eq:XYMatrices}
	X_{k,j} &= c^2 \prtq{\Tr \prtg{F_j \prt{\cst }A_k^\dagger} + \hc},\nonumber \\
	Y_{k,j} &= \frac{c^2}{2 i} \prtq{\Tr \prtg{F_j \prt{\cst} A_k^\dagger} -\hc}.
\end{align}
The matrix $X$ is Hermitian and positive semi-definite.
Then, there exists a non-unique $M \times M$ matrix $\Lambda$ such that $X=\Lambda \Lambda^\dagger$.

The second order dynamics is given by~\cite{ThesisAzouit2017}
\begin{equation}
	\label{eq:SecondOrderResume}
	\dot{\rho}_B = \epsilon \Lb_{S,1} (\rho_B) + \epsilon^2 \Lb_{S,2} (\rho_B),
\end{equation}  
where
\begin{align}\label{eq:SecondOrderResume2}
	\Lb_{S,1} (\rho_B) &= -i \sum_{k=1}^M \comm{\Tr_A \prtg{A_k \cst} B_k^\dagger}{\rho_B } + \Lb_B (\rho_B),\nonumber \\
	\Lb_{S,2} (\rho_B) &= -i \comm{\sum_{k,j=1}^M Y_{k,j}B_k B_j^\dagger}{\rho_B}
	+\sum_{p=1}^M \D_{L_p} (\rho_B),\nonumber \\
	L_p  &= \sum_{j=1}^M \Lambda^*_{j,p} B^\dagger_j.
\end{align}

\section{Adiabatic elimination of the harmonic oscillator\label{APPSec:AdiabaticEliminationApplied}}

Here, we apply the general method described in the previous section to the one-photon (1ph) and the 2ph models considered in the main text. There, system $A$ consists of an harmonic oscillator (HO) while system $B$ is an ensemble of TLSs. In particular, the starting point for this application is the Eq.~(2) of the main text which describes the global dynamics of 
the two models and is given in the interaction picture.
In the Schr\"{o}dinger picture, the two models are described by the equation
\begin{equation}\label{eq:FullModelSP}
	\dot{\rho}_{G,S}=  - \frac{i}{\hbar} \comm{H_l}{\rho_{G,S}} + \Lb_{\textup{HO},S}(\rho_{G,S})+\Lb_{Q,S} (\rho_{G,S}) ,
\end{equation}
where $\rho_{G,S}$ is the global density matrix in the Schr\"{o}dinger picture,  $l=1$ for the 1ph case and $l=2$ for the 2ph one, $H_l$  is given in Eq.~(1) of the main text, and
\begin{align}	
	\Lb_{\textup{HO},S} \prt{\bullet}= &   -i \comm{\prt{ \beta^* a e^{i \omega t} + \beta \ad e^{-i \omega t} }}{\bullet} \nonumber \\
	&+k\prtq{\prt{1+\nwt}\D_{a}\prt{\bullet}+\nwt\D_{\ad}\prt{\bullet}}, \nonumber  \\
	\Lb_{Q,S} (\bullet)=&  \sum_{i=1}^{N} \left[\gamloc\prt{1+\nlwt}\D_{\sigma_-^{(i)}}\prt{\bullet} \right. \nonumber \\
	&\left.+\prt{\gamloc\nlwt+P}\D_{\sigma_+^{(i)}}\prt{\bullet}\right].
\end{align}
The passage from the Schr\"{o}dinger picture to the interaction one is, indeed, necessary to apply the adiabatic elimination method since the bare dynamics of both the TLSs and the harmonic oscillator is much faster than all other dynamics so that to separate the system into a fast and a slow part is not possible. Moreover, neglecting the interaction between the two subsystems, a steady state of the HO does not exist because of the time dependent part in the Hamiltonian describing the action of the coherent driving. The unitary operator used to move from the Schr\"{o}dinger picture to the interaction one [i.e., from Eq.~\eqref{eq:FullModelSP}  to Eq.~(2) of the main text, being $\rho_{G}=U_l^\dagger (t)\rho_{G,S}U_l (t)$] is
\begin{equation}
	U_l (t) = e^{-\frac{i}{\hbar} H_{0,l} t}, 
\end{equation}
where
\begin{equation}
	H_{0,l} =  \hbar \omega \ad a+\frac{l \hbar\omega}{2} J_z.
\end{equation}
In this section, we denote the reduced density matrix of the HO with $\rho_A$ in order to maintain the notation of the preceding section, while we call the reduced density matrix of the TLSs $\rho$, without suffixes, in order to have the same notation of the main text.

As explained in the preceding section, we first need to obtain the steady state of the HO when it does not interact with the TLSs.
This steady state is equal to~\cite{ThesisAzouit2017}
\begin{equation}\label{eq:steadystatecavity0}
	\cst = D(\alpha) \rth D(-\alpha),
\end{equation}	
where $ D(\alpha)=e^{\alpha a^\dagger - \alpha^* a}$,  $\alpha= - \frac{2 i \beta}{k}$, and
\begin{equation}\label{eq:steadystatecavity1}
	\rth = \frac{1}{1+\nwt}\sum_{n=0}^\infty \prt{\frac{\nwt}{1+\nwt}}^n \dyad{n}.
\end{equation}
In other words, $\cst$ is a thermal coherent state obtained by applying to the thermal state with average energy $\hbar \omega \nwt$ the displacement operator corresponding to the coherent state in which the HO would be at zero temperature.

We choose to use as perturbative parameter $\epsilon$ the quantity $g/k$ so that $\epsilon = g/k$.
With this choice, we can write
\begin{equation}
	\intH = \hbar k \prtq{ \an J_+ + \adn J_-},
\end{equation}
where $c= k$.
Moreover, $\Lb_Q = \epsilon \Lb_B$.
Therefore, at first order in $g/k$ we get [see Eq.~\eqref{eq: first order reduced dynamics2} with $A_1 = \an$, $A_2 = \adn$, $B^\dagger_1 = J_+$, and $B^\dagger_2 = J_-$]
\begin{equation}\label{eq: first order reduced dynamics our model}
	\epsilon \Lb_{S,1} (\rho) = -i g \comm{\alpha^l J_+ + \prt{\alpha^*}^l J_-}{\rho}+\Lb_Q(\rho),
\end{equation}
since [using $a D(\alpha)=\alpha D(\alpha) + D(\alpha) a$]
\begin{align}
	\Tr \prtg{\an \cst} 
	&= \Tr \prtg{\alpha a^{l-1} \cst + D(\alpha) a \rth D(-\alpha)} \nonumber \\
	&=\Tr \prtg{\alpha a^{l-1} \cst}+\Tr \prtg{a \rth} \nonumber\\
	&=\Tr \prtg{\alpha^2 a^{l-2} \cst}+\alpha \Tr \prtg{a \rth} \nonumber\\
	&= \vdots\nonumber\\
	&= \alpha^l, \nonumber\\
	\Tr \prtg{\adn \cst} 
	&= \prt{\Tr \prtg{\an \cst}}^*=\prt{\alpha^*}^l.
\end{align}

To obtain the second order dynamics we need the superoperators $F_k$.
In this case, there are only two of them:
\begin{align}
	F_1 \prt{\cst} &= \int_0^{\infty}e^{t\Lb_A}\prtq{\prt{\an-\alpha^l}\cst}\dd{t},\nonumber \\
	F_2 \prt{\cst} &= \int_0^{\infty}e^{t\Lb_A}\prtq{\prt{\adn-\prt{\alpha^*}^l}\cst}\dd{t}.
\end{align}
In order to calculate the matrix elements $X_{i,j}$ and $Y_{i,j}$ of Eq.~\eqref{eq:XYMatrices} we just need to compute terms like $\Tr \prtg{F_1 \cst \adn}$ so that finding the explicit form of the operators $F_k$ is not necessary.
Since $\Tr \prtg{F_i \prt{\cst}} = 0$, one can, for example, write
\begin{align}
	\Tr &\prtg{F_1 \cst \adn}
	= \Tr \prtg{\adn \int_0^{\infty}e^{t\Lb_A}\prtq{(\an-\alpha^l)\cst}\dd{t}}\nonumber \\
	&= \Tr \prtg{\prtq{\adn-\alfsn} \int_0^{\infty}e^{t\Lb_A}\prtq{(\an-\alpha^l)\cst}\dd{t}}\nonumber 
	\\
	&= \Tr \prtg{\prtq{(\an-\alpha^l)\cst}\int_0^{\infty}e^{t\Lb^\dagger_A} \prtq{\adn-\alfsn}\dd{t}},
\end{align}
where $\Lb^\dagger_A$ is the adjoint Lindblad operator~\cite{BookBreuer2007,ThesisAzouit2017}.
In the same way, one obtains the following quantities
\begin{widetext}
	\begin{align}
		\label{eq:Fterms}
		\Tr \prtg{F_1 \cst \adn}
		&= \Tr \prtg{\prtq{(\an-\alpha^l)\cst}\int_0^{\infty}e^{t\Lb^\dagger_A} \prtq{\adn-\alfsn}\dd{t}},\nonumber \\
		\Tr \prtg{F_1 \cst \an}
		&= \Tr \prtg{\prtq{(\an-\alpha^l)\cst}\int_0^{\infty}e^{t\Lb^\dagger_A} \prt{\an-\alpha^l}\dd{t}},\nonumber \\
		\Tr \prtg{F_2 \cst \adn}
		&= \Tr \prtg{\prtq{\prt{\adn-\alfsn}\cst}\int_0^{\infty}e^{t\Lb^\dagger_A} \prtq{\adn-\alfsn}\dd{t}},\nonumber \\
		\Tr \prtg{F_2 \cst \an}
		&= \Tr \prtg{\prtq{\prt{\adn-\alfsn}\cst}\int_0^{\infty}e^{t\Lb^\dagger_A} \prt{\an-\alpha^l}\dd{t}}.
	\end{align}
\end{widetext}

The above formulas (and the resulting master equation) are valid for any $l$ but, from now on, we will deal explicitly with the $l=1,2$ cases because otherwise calculations become needlessly cumbersome.
The result obtained in the case $l=1$ is already known in literature \cite{ThesisAzouit2017}.
Nevertheless, we think that reporting here its derivation with this method can be helpful.

\subsection{The 1ph case}

For $l=1$ it is possible to write $e^{t\Lb_A^\dagger}a= f_0(t) +f_1(t)a$, with $f_0 (0) =0$ and $f_1(0)=1$. The adjoint master equation for the operator $e^{t\Lb_A^\dagger}a$ reads~\cite{ThesisAzouit2017,BookBreuer2007}
\begin{equation}
	\dv{t} f_0 (t)+ \dv{t} f_1 (t) a  = -\frac{k}{2} f_1 (t) a + \alpha\frac{k}{2}f_1 (t),
\end{equation}
whose solution,  $f_0 (t) =\alpha \prt{1-e^{-\frac{k}{2}t}}$ and $f_1 (t) = e^{-\frac{k}{2}t}$, implies
\begin{equation}
	\label{eq:OnePhotonHeisenberg}
	e^{t\Lb_A^\dagger}\prt{a-\alpha}=e^{-\frac{k}{2}t}\prt{a-\alpha}.
\end{equation}
We recall that $e^{t\Lb_A^\dagger} r = r$, where $r$ is a constant.
Analogously, we can write $e^{t\Lb_A^\dagger} \ad= \tilde{f}_0(t)+ \tilde{f}_1(t)\ad $ and solve the associated differential equations. The solutions are equal to the ones for $f_0 (t)$ and $f_1 (t)$ with the substitution $\alpha \rightarrow \alpha^*$. It follows that
\begin{equation}
	\label{eq:OnePhotonHeisenberg2}
	e^{t\Lb_A^\dagger}\prt{\ad-\alpha^*}=e^{-\frac{k}{2}t}\prt{\ad-\alpha^*}.
\end{equation}

Now, we can calculate the elements of the $X$ and $Y$ matrices.
Using Eqs.~\eqref{eq:OnePhotonHeisenberg} and \eqref{eq:OnePhotonHeisenberg2} in Eq.~\eqref{eq:Fterms}, we obtain for $l=1$ (hereafter $\nuno=\nwt = [e^{\hbar \omega /(k_B T)}-1]^{-1}$)
\begin{align}
	\label{eq:FtermsOnePhoton}
	&\Tr \prtg{F_1 \cst \ad}
	= \frac{2}{k} \nuno, &\Tr& \prtg{F_1 \cst a} =0,\nonumber \\
	&\Tr \prtg{F_2 \cst \ad} = 0,  &\Tr& \prtg{F_2 \cst a} =\frac{2}{k}\prt{1+\nuno}.
\end{align}
By inserting Eq.~\eqref{eq:FtermsOnePhoton} in Eq.~\eqref{eq:XYMatrices}, we then have
\begin{equation}
	X = 4 k\mqty(\dmat[0]{\nuno,1+\nuno}),\quad Y = \mqty(\dmat[0]{0,0}).
\end{equation}
We can set $\Lambda=\sqrt{X}$ thus obtaining in Eq.~\eqref{eq:SecondOrderResume2}:
\begin{align}
	L_1 
	&= \Lambda^*_{1,1} J_+ + \Lambda^*_{2,1} J_- = \sqrt{4 k \nuno} J_+,\nonumber \\
	L_2
	&= \Lambda^*_{1,2} J_+ + \Lambda^*_{2,2} J_- = \sqrt{4 k \prt{1+\nuno}} J_-.
\end{align}
Eventually, using Eqs.~\eqref{eq:SecondOrderResume} and \eqref{eq: first order reduced dynamics our model}, we obtain as equation for the second order dynamics of the TLSs
\begin{align}
	\label{eq:MasterEquation1photon}
	\dot{\rho} = &-i g \comm{\alpha J_+ + \alpha^* J_-}{\rho} + \Lb_Q (\rho) + \nonumber \\
	&+ \gamma_1 \prtq{\nuno \D_{J_+} (\rho) + \prt{1+\nuno } \D_{J_-} (\rho) } ,
\end{align}
where $\gamma_1 = 4 g^2/k$.

\subsection{The 2ph case}

The derivation of the reduced dynamics for $l=2$ proceeds analogously to the $l=1$ case, but it is more involved.
We can write $e^{t \Lb_A^\dagger }a^2 = h_0 (t)+ h_1(t) a + h_2(t) a^2  $, with $h_0 (0) = h_1 (0) =0$ and $h_2 (0) =1$.
The adjoint master equation for the operator $e^{t \Lb_A^\dagger }a^2$ reads
\begin{multline}
	\dv{t}  h_0 (t)+ \dv{t} h_1(t) a  + \dv{t} h_2(t) a^2=  \frac{\alpha k}{2} h_1 (t)\\  
	+ \prtq{\alpha k h_2 (t) - \frac{k}{2} h_1 (t)}a -k h_2 (t) a^2,
\end{multline}
whose solution, $h_0 (t) = \alpha^2 e^{-kt}\prt{e^{\frac{k}{2}t}-1}^2$, $h_1 (t) = 2 \alpha e^{-kt}\prt{e^{\frac{k}{2}t}-1}$, and $h_2 (t) = e^{-kt}$, implies
\begin{equation}
	\label{eq:TwoPhotonHeisenberg}
	e^{t\Lb_A^\dagger}\prt{a^2 - \alpha^2} = 
	e^{-kt}\prt{a -\alpha}^2 +e^{-\frac{k}{2}t}\prtq{2\alpha\prt{a-\alpha}}.
\end{equation}
Analogously, we can write  $e^{t \Lb_A^\dagger } (\ad)^2 = \tilde{h}_0 (t)+ \tilde{h}_1(t) \ad+$ $ \tilde{h}_2(t) (\ad)^2 $ and solve the associated differential equations. The solutions are equal to the ones for $h_0 (t)$, $h_1 (t)$, and $h_2 (t)$ with the substitution $\alpha \rightarrow \alpha^*$:
\begin{align}
	\label{eq:TwoPhotonHeisenberg2}
	e^{t\Lb_A^\dagger}\prtq{\add - \prt{\alpha^*}^2} =\:\: & 
	e^{-kt}\prt{\ad -\alpha^*}^2 \nonumber \\&+e^{-\frac{k}{2}t}\prtq{2 \alpha^* \prt{\ad-\alpha^*}}.
\end{align}

Now, we can calculate the elements of the $X$ and $Y$ matrices.
After straightforward but lengthy calculations, using Eqs.~\eqref{eq:TwoPhotonHeisenberg} and \eqref{eq:TwoPhotonHeisenberg2} in Eq.~\eqref{eq:Fterms} and the following equalities
\begin{align}
	D(-\alpha)(a-\alpha) &= a D(-\alpha), \nonumber \\  D(-\alpha)(\ad-\alpha^*) &= \ad D(-\alpha), \nonumber \\
	\prt{a^2 - \alpha^2}D(\alpha) &= D(\alpha)\prt{a + 2 \alpha}a, \nonumber \\
	\prtq{\add - \prt{\alpha^*}^2}D(\alpha) &= D(\alpha)\prt{\ad + 2 \alpha^*}\ad, \nonumber\\
	\Tr \prtg{\add a^2 \rth} &=  2 (\nuno)^2, \nonumber \\
	\Tr \prtg{a^2 \add \rth} &=  2 (1+\nuno)^2,
\end{align}
we obtain, for $l=2$,
\begin{align}
	\label{eq:FtermsTwoPhoton}
	\Tr \prtg{F_1 \cst \add} &= \frac{2}{k} \prtq{\prt{\nuno}^2 + 4 \alphadue \nuno },\nonumber \\
	\Tr \prtg{F_2 \cst a^2}  &= \frac{2}{k} \prtq{\prt{1+\nuno}^2 + 4\alphadue \prt{1+\nuno}},\nonumber \\
	\Tr \prtg{F_1 \cst a^2}  &= \Tr \prtg{F_2 \cst \add} = 0.
\end{align}
Then, the $X$ and $Y$ matrices are easily obtained by inserting Eq.~\eqref{eq:FtermsTwoPhoton} in Eq.~\eqref{eq:XYMatrices}:
\begin{align}
	X &= 4 k\mqty(\dmat[0]{\prt{\nuno}^2 + 4 \alphadue \nuno ,\prt{1+\nuno}^2 + 4\alphadue \prt{1+\nuno}}), \nonumber \\ 
	Y &= \mqty(\dmat[0]{0,0}).
\end{align}
Eventually, using Eq.~\eqref{eq:SecondOrderResume} and setting $\Lambda=\sqrt{X}$ as in the 1ph case, the second order dynamics of the TLSs reads
\begin{align}
	\dot{\rho} 
	&= -i g \comm{\alpha^2 J_+ + \prt{\alpha^*}^2 J_-}{\rho} + \Lb_Q (\rho)\nonumber \\
	&\quad + \gamma_1\Bigg[\prt{4 \alphadue \nuno + \prt{\nuno}^2}\D_{J_+} (\rho)\nonumber \\
	&\quad + \prt{4\alphadue \prt{1+\nuno} + \prt{1+\nuno}^2} \D_{J_-} (\rho) \Bigg].
\end{align}

The above equation can be rewritten by operating the following substitutions
\begin{align}
	\gamma_1 \prtq{4 \alphadue \nuno + \prt{\nuno}^2} &\rightarrow \gamma_2 \ndue \nonumber \\
	\gamma_1 \prtq{4\alphadue \prt{1+\nuno} + \prt{1+\nuno}^2} &\rightarrow \gamma_2 \prt{1+\ndue},
\end{align}
where 
\begin{equation}
	\ndue = \nuno \frac{\nuno + 4\alphadue}{1+2\nuno + 4\alphadue},\
	\gamma_2 = \gamma_1 \prt{1 + 2 \nuno + 4 \alphadue}.
\end{equation}
In this way, the 2ph model reduced master equation becomes
\begin{align}\label{eq:ReducedMasterEquations}
	\dot{\rho} 
	=& -i g \comm{\alpha^2 J_+ + \prt{\alpha^*}^2 J_-}{\rho} + \Lb_Q (\rho)\nonumber \\ &+\gamma_2\prtq{\ndue \D_{J_+}+ \prt{1+\ndue} \D_{J_-}}\rho.
\end{align}
Notice that the quantity $\ndue$ can also be written as
\begin{equation}
	\label{eq:ndueEquivalent}
	\ndue = \ndwt + \frac{4 \nuno \alphadue}{1 + 2 \nuno + 4 \alphadue} \frac{1 + \nuno}{1 + 2 \nuno},
\end{equation}
which is another way to display what is shown in Eq.~(6) of the main text, i.e., the fact that the effective temperature of the collective bath as seen by the TLSs is higher than the actual temperature $T$ due to the action of the external coherent field on the harmonic oscillator.

\subsection{Validity of the adiabatic elimination}

We finally comment on the validity of the adiabatic elimination approximation which, in our setting, relies on the much higher rate of losses of the HO compared to its exchanges with the TLSs and requires stronger conditions than just $g \ll k$.
For example, for the 1ph coupling with one TLS,  if the HO is in a Fock state with $\tilde{n}$ excitations and the TLS is in the ground state, the \enquote{Rabi oscillations}  have angular frequency $g\sqrt{\tilde{n}}$, leading to  the condition $g\sqrt{\tilde{n}} \ll k$. In the 2ph case, the same reasoning leads to $g\sqrt{\tilde{n}(\tilde{n}-1)} \ll k$. In our dynamics, at order zero [see Eqs.~\eqref{eq:steadystatecavity0} and \eqref{eq:steadystatecavity1}] the HO is in a thermal coherent state with an average number of excitations $\bar{n}=\abs{\alpha}^2+\nwt$. Then, we can roughly estimate the validity of the adiabatic elimination by using this value for $\tilde{n}$ in the above conditions. In general, we expect the approximation to not work properly also when $P \gtrsim k$ since in this case the TLSs emission would compete with the HO losses.

\section{Mathematical mapping of the two models\label{APPSec:MathematicalMapping}}

The comparison of Eq.~\eqref{eq:MasterEquation1photon} and Eq.~\eqref{eq:ReducedMasterEquations} shows that the master equation describing the dynamics of the TLSs can be cast in the same form for both models.
For this reason, a mathematical mapping between the two models is obtainable.
In particular, given the physical parameters in the 2ph model
\begin{equation}
	g,\
	\beta,\
	k,\
	\omega,\
	T,\
	P,\
	\gamloc,\
\end{equation}
contained in the effective parameters $\alpha,\ \nuno, \ndue ,\ndwt$,
the dynamics to which they give rise can be obtained in the 1ph model by different proper choices of the physical parameters.
Denoting with an apex the effective quantities for this \enquote{simulation} in the 1ph model we get
\begin{align}
	\frac{g'}{\sqrt{k'}} &=  \frac{g \sqrt{1+2\nuno + 4\alphadue}}{\sqrt{k}},\nonumber\\
	\alpha' \sqrt{k'} &= \frac{\alpha^2 \sqrt{k}}{\sqrt{1+2\nuno + 4\alphadue}},\nonumber\\
	n_1' &\equiv \overline{n}_{\omega', T'} = \ndue, \nonumber\\
	P' &= P - \prt{\ndue - \ndwt} \frac{\gamloc}{1+\ndue}, \nonumber\\
	\gamloc' &= \gamloc \frac{1+\ndwt}{1+\ndue}.
\end{align}

We remark, however, that this mapping between the two models is a mathematical mapping and that there are situations in which the dynamics obtained in one model is not obtainable in the other one.
For example,  when the incoherent pumping is absent in the 2ph model, since $\ndue > \ndwt$ [see Eq.~\eqref{eq:ndueEquivalent}], in the 1ph model we could {need} $P' < 0$ to simulate the 2ph model, and this does not correspond to the case of an incoherent pumping term.
Indeed, the physical reason for this incompatibility is the temperature-modifying effect of the coherent driving taking place only in the 2ph model [see Eq.~(6) of the main text].

\section{Steady states\label{APPSec:SteadyStates}}

In this section, we report some details regarding the steady states of the TLS dynamics in the one- and two- TLS cases.

\subsection{One-TLS case\label{APPSubSec:OneTLSCase}}

In the case when system $B$ consists of just one TLS, it is described by the master equation
\begin{align}
	\label{eq:OneAtomReducedMasterEquation}
	\dot{\rho} =&	-i g \comm{\alpha^l \sigma_+ + \alpha^{*l}\sigma_-}{\rho} \nonumber \\
	&+\prtq{\Gamma_l^{(-)}\D_{\sigma_-}\prt{\rho} +\Gamma_l^{(+)}\D_{\sigma_+}\prt{\rho}},
\end{align}
where $\Gamma_l^{(-)} = \gamloc (1+\nlwt) + \gamma_l (1+n_l)$ and $\Gamma_l^{(+)} = \gamloc \nlwt  + \gamma_l n_l + P$.
The density matrix elements of the steady state $\rho^{\textup{st}}$ are found to be
\begin{align}
	\rho^{\textup{st}}_{ee} &= \frac{4 g^2 \abs{\alpha}^{2l} + \Gamma_l^{(+)}\prt{\Gamma_l^{(-)} + \Gamma_l^{(+)}}}{8 g^2 \abs{\alpha}^{2l} + \prt{\Gamma_l^{(-)} + \Gamma_l^{(+)}}^2},\nonumber \\
	\rho^{\textup{st}}_{eg} &= \frac{ 2 i g \alpha^l\prt{{\Gamma_l^{(+)}-\Gamma_l^{(-)}}}}{8 g^2 \abs{\alpha}^{2l} + \prt{\Gamma_l^{(-)} + \Gamma_l^{(+)}}^2},
\end{align}
being for any $\rho$, $\rho_{gg}=1-\rho_{ee}$ and $\rho_{ge}=\rho_{eg}^*$.

An interesting limit case is obtained when $\abs{\alpha}$ is high enough that every term not containing it can be safely neglected.
In the 1ph case, the result of this operation is
\begin{equation}
	\rho^{\textup{st}}_{ee} = \frac{1}{2}, \quad \rho^{\textup{st}}_{eg} = 0.
\end{equation}
In the 2ph case, in this limit we have $\Gamma_2^{(-)} \simeq 16 (1+\nuno ) \abs{\alpha}^2 g^2/k$ and $\Gamma_2^{(+)} \simeq 16 \nuno \abs{\alpha}^2 g^2/k$ so that
\begin{align}
	\rho^{\textup{st}}_{ee} &= \frac{1 + 64 \nuno \prt{ 1+ 2 \nuno } \prt{g/k}^2}{2+64 \prt{1+ 2 \nuno}^2 \prt{g/k}^2},\nonumber \\
	\rho^{\textup{st}}_{eg} &= e^{i\prt{2\phi - \frac{\pi}{2}}}\frac{4 g/k}{ 1 + 32 \prt{1+2 \nuno}^2 \prt{g/k}^2},
\end{align}
where we have used the notation $\alpha=\abs{\alpha}e^{i\phi}$.
The case represented in Fig.~2 of the main text is the zero-temperature one, for which the above formulas become
\begin{equation}
	\rho^{\textup{st}}_{ee} = \frac{1}{2+64(g/k)^2},\quad
	\rho^{\textup{st}}_{eg}= e^{i\prt{2\phi - \frac{\pi}{2}}}\frac{4 g/k}{ 1 + 32 (g/k)^2}.
\end{equation}
In this case, the maximum of $\abs{\rho^{\textup{st}}_{eg}}$ is obtained for $g/k= 1/(4\sqrt{2})\simeq 0.177$. However, for this value of $g/k$ we are not anymore in the bad-cavity limit.

\subsection{Two-TLS case\label{APPSubSec:TwoTLSCase}}

The steady state of two TLSs can be found analytically (we have done it using MATHEMATICA) but its form is very cumbersome and, therefore, we do not report it here.
Here, we report the analytical form of $\Jcorr$ in the case $\alpha=0$, analyzed in the maintext:
\begin{widetext}
	\begin{equation}\label{eq:Jcorr}
		\Jcorr=\frac{P \gamma_l (1+R_l)(P-\gamma_l-\gamloc)}{\prt{P+\gamloc R_l}^3 + 3 \gamma_l R_l \prt{P + \gamloc R_l}^2 +\gamma_l^2 \prtq{2\gamloc R_l^3 + P \prt{1 + R_l + 2 R_l^2}}},
	\end{equation}
\end{widetext}
where $R_l=1+2 n_l$.
The sign of $\Jcorr$ in Eq.~\eqref{eq:Jcorr} depends only on $P-\gamma_l-\gamloc$, which is temperature dependent only in the 2ph case.
Our numerical simulations of the effective model indicate that this is true for any number of TLSs.
We have checked this up to six of them.


\begin{thebibliography}{67}%
	\makeatletter
	\providecommand \@ifxundefined [1]{%
		\@ifx{#1\undefined}
	}%
	\providecommand \@ifnum [1]{%
		\ifnum #1\expandafter \@firstoftwo
		\else \expandafter \@secondoftwo
		\fi
	}%
	\providecommand \@ifx [1]{%
		\ifx #1\expandafter \@firstoftwo
		\else \expandafter \@secondoftwo
		\fi
	}%
	\providecommand \natexlab [1]{#1}%
	\providecommand \enquote  [1]{``#1''}%
	\providecommand \bibnamefont  [1]{#1}%
	\providecommand \bibfnamefont [1]{#1}%
	\providecommand \citenamefont [1]{#1}%
	\providecommand \href@noop [0]{\@secondoftwo}%
	\providecommand \href [0]{\begingroup \@sanitize@url \@href}%
	\providecommand \@href[1]{\@@startlink{#1}\@@href}%
	\providecommand \@@href[1]{\endgroup#1\@@endlink}%
	\providecommand \@sanitize@url [0]{\catcode `\\12\catcode `\$12\catcode
		`\&12\catcode `\#12\catcode `\^12\catcode `\_12\catcode `\%12\relax}%
	\providecommand \@@startlink[1]{}%
	\providecommand \@@endlink[0]{}%
	\providecommand \url  [0]{\begingroup\@sanitize@url \@url }%
	\providecommand \@url [1]{\endgroup\@href {#1}{\urlprefix }}%
	\providecommand \urlprefix  [0]{URL }%
	\providecommand \Eprint [0]{\href }%
	\providecommand \doibase [0]{https://doi.org/}%
	\providecommand \selectlanguage [0]{\@gobble}%
	\providecommand \bibinfo  [0]{\@secondoftwo}%
	\providecommand \bibfield  [0]{\@secondoftwo}%
	\providecommand \translation [1]{[#1]}%
	\providecommand \BibitemOpen [0]{}%
	\providecommand \bibitemStop [0]{}%
	\providecommand \bibitemNoStop [0]{.\EOS\space}%
	\providecommand \EOS [0]{\spacefactor3000\relax}%
	\providecommand \BibitemShut  [1]{\csname bibitem#1\endcsname}%
	\let\auto@bib@innerbib\@empty
	\bibitem [{\citenamefont {Haroche}\ and\ \citenamefont
		{Raimond}(2006)}]{BookHaroche2006}%
	\BibitemOpen
	\bibfield  {author} {\bibinfo {author} {\bibfnamefont {S.}~\bibnamefont
			{Haroche}}\ and\ \bibinfo {author} {\bibfnamefont {J.-M.}\ \bibnamefont
			{Raimond}},\ }\href
	{https://doi.org/10.1093/acprof:oso/9780198509141.001.0001} {\emph {\bibinfo
			{title} {Exploring The Quantum: Atoms, Cavities, and Photons}}}\ (\bibinfo
	{publisher} {Oxford University Press},\ \bibinfo {year} {2006})\BibitemShut
	{NoStop}%
	\bibitem [{\citenamefont {Kirton}\ \emph {et~al.}(2019)\citenamefont {Kirton},
		\citenamefont {Roses}, \citenamefont {Keeling},\ and\ \citenamefont
		{Dalla~Torre}}]{KirtonReview}%
	\BibitemOpen
	\bibfield  {author} {\bibinfo {author} {\bibfnamefont {P.}~\bibnamefont
			{Kirton}}, \bibinfo {author} {\bibfnamefont {M.~M.}\ \bibnamefont {Roses}},
		\bibinfo {author} {\bibfnamefont {J.}~\bibnamefont {Keeling}},\ and\ \bibinfo
		{author} {\bibfnamefont {E.~G.}\ \bibnamefont {Dalla~Torre}},\ }\bibfield
	{title} {\bibinfo {title} {Introduction to the {Dicke} {Model}: From
			{Equilibrium} to {Nonequilibrium}, and vice versa},\ }\href
	{https://doi.org/10.1002/qute.201800043} {\bibfield  {journal} {\bibinfo
			{journal} {Adv. Quantum Technol.}\ }\textbf {\bibinfo {volume} {2}},\
		\bibinfo {pages} {1800043} (\bibinfo {year} {2019})}\BibitemShut {NoStop}%
	\bibitem [{\citenamefont {Benedict}(1996)}]{benedict2018super}%
	\BibitemOpen
	\bibfield  {author} {\bibinfo {author} {\bibfnamefont {M.~G.}\ \bibnamefont
			{Benedict}},\ }\href {https://doi.org/10.1201/9780203737880} {\emph {\bibinfo
			{title} {Super-Radiance: Multiatomic Coherent Emission}}}\ (\bibinfo
	{publisher} {CRC, Boca Raton, FL},\ \bibinfo {year} {1996})\BibitemShut {NoStop}%
	\bibitem [{\citenamefont {Gross}\ and\ \citenamefont
		{Haroche}(1982)}]{GROSS1982301}%
	\BibitemOpen
	\bibfield  {author} {\bibinfo {author} {\bibfnamefont {M.}~\bibnamefont
			{Gross}}\ and\ \bibinfo {author} {\bibfnamefont {S.}~\bibnamefont
			{Haroche}},\ }\bibfield  {title} {\bibinfo {title} {Superradiance: An essay
			on the theory of collective spontaneous emission},\ }\href
	{https://doi.org/https://doi.org/10.1016/0370-1573(82)90102-8} {\bibfield
		{journal} {\bibinfo  {journal} {Phys. Rep.}\ }\textbf {\bibinfo {volume}
			{93}},\ \bibinfo {pages} {301} (\bibinfo {year} {1982})}\BibitemShut
	{NoStop}%
	\bibitem [{\citenamefont {Meiser}\ \emph {et~al.}(2009)\citenamefont {Meiser},
		\citenamefont {Ye}, \citenamefont {Carlson},\ and\ \citenamefont
		{Holland}}]{Meiser2009}%
	\BibitemOpen
	\bibfield  {author} {\bibinfo {author} {\bibfnamefont {D.}~\bibnamefont
			{Meiser}}, \bibinfo {author} {\bibfnamefont {J.}~\bibnamefont {Ye}}, \bibinfo
		{author} {\bibfnamefont {D.~R.}\ \bibnamefont {Carlson}},\ and\ \bibinfo
		{author} {\bibfnamefont {M.~J.}\ \bibnamefont {Holland}},\ }\bibfield
	{title} {\bibinfo {title} {Prospects for a {Millihertz-Linewidth} {Laser}},\
	}\href {https://doi.org/10.1103/PhysRevLett.102.163601} {\bibfield  {journal}
		{\bibinfo  {journal} {Phys. Rev. Lett.}\ }\textbf {\bibinfo {volume} {102}},\
		\bibinfo {pages} {163601} (\bibinfo {year} {2009})}\BibitemShut {NoStop}%
	\bibitem [{\citenamefont {Meiser}\ and\ \citenamefont
		{Holland}(2010{\natexlab{a}})}]{Meiser2010I}%
	\BibitemOpen
	\bibfield  {author} {\bibinfo {author} {\bibfnamefont {D.}~\bibnamefont
			{Meiser}}\ and\ \bibinfo {author} {\bibfnamefont {M.~J.}\ \bibnamefont
			{Holland}},\ }\bibfield  {title} {\bibinfo {title} {Steady-state
			superradiance with alkaline-earth-metal atoms},\ }\href
	{https://doi.org/10.1103/PhysRevA.81.033847} {\bibfield  {journal} {\bibinfo
			{journal} {Phys. Rev. A}\ }\textbf {\bibinfo {volume} {81}},\ \bibinfo
		{pages} {033847} (\bibinfo {year} {2010}{\natexlab{a}})}\BibitemShut
	{NoStop}%
	\bibitem [{\citenamefont {Meiser}\ and\ \citenamefont
		{Holland}(2010{\natexlab{b}})}]{Meiser2010II}%
	\BibitemOpen
	\bibfield  {author} {\bibinfo {author} {\bibfnamefont {D.}~\bibnamefont
			{Meiser}}\ and\ \bibinfo {author} {\bibfnamefont {M.~J.}\ \bibnamefont
			{Holland}},\ }\bibfield  {title} {\bibinfo {title} {Intensity fluctuations in
			steady-state superradiance},\ }\href
	{https://doi.org/10.1103/PhysRevA.81.063827} {\bibfield  {journal} {\bibinfo
			{journal} {Phys. Rev. A}\ }\textbf {\bibinfo {volume} {81}},\ \bibinfo
		{pages} {063827} (\bibinfo {year} {2010}{\natexlab{b}})}\BibitemShut
	{NoStop}%
	\bibitem [{\citenamefont {Auff{\`{e}}ves}\ \emph {et~al.}(2011)\citenamefont
		{Auff{\`{e}}ves}, \citenamefont {Gerace}, \citenamefont {Portolan},
		\citenamefont {Drezet},\ and\ \citenamefont {Santos}}]{Auffeves2011}%
	\BibitemOpen
	\bibfield  {author} {\bibinfo {author} {\bibfnamefont {A.}~\bibnamefont
			{Auff{\`{e}}ves}}, \bibinfo {author} {\bibfnamefont {D.}~\bibnamefont
			{Gerace}}, \bibinfo {author} {\bibfnamefont {S.}~\bibnamefont {Portolan}},
		\bibinfo {author} {\bibfnamefont {A.}~\bibnamefont {Drezet}},\ and\ \bibinfo
		{author} {\bibfnamefont {M.~F.}\ \bibnamefont {Santos}},\ }\bibfield  {title}
	{\bibinfo {title} {Few emitters in a cavity: From cooperative emission to
			individualization},\ }\href {https://doi.org/10.1088/1367-2630/13/9/093020}
	{\bibfield  {journal} {\bibinfo  {journal} {New J. Phys.}\ }\textbf {\bibinfo
			{volume} {13}},\ \bibinfo {pages} {093020} (\bibinfo {year}
		{2011})}\BibitemShut {NoStop}%
	\bibitem [{\citenamefont {Shammah}\ \emph {et~al.}(2017)\citenamefont
		{Shammah}, \citenamefont {Lambert}, \citenamefont {Nori},\ and\ \citenamefont
		{De~Liberato}}]{PhysRevA.96.023863}%
	\BibitemOpen
	\bibfield  {author} {\bibinfo {author} {\bibfnamefont {N.}~\bibnamefont
			{Shammah}}, \bibinfo {author} {\bibfnamefont {N.}~\bibnamefont {Lambert}},
		\bibinfo {author} {\bibfnamefont {F.}~\bibnamefont {Nori}},\ and\ \bibinfo
		{author} {\bibfnamefont {S.}~\bibnamefont {De~Liberato}},\ }\bibfield
	{title} {\bibinfo {title} {Superradiance with local phase-breaking effects},\
	}\href {https://doi.org/10.1103/PhysRevA.96.023863} {\bibfield  {journal}
		{\bibinfo  {journal} {Phys. Rev. A}\ }\textbf {\bibinfo {volume} {96}},\
		\bibinfo {pages} {023863} (\bibinfo {year} {2017})}\BibitemShut {NoStop}%
	\bibitem [{\citenamefont {Kirton}\ and\ \citenamefont
		{Keeling}(2018)}]{kirton2018}%
	\BibitemOpen
	\bibfield  {author} {\bibinfo {author} {\bibfnamefont {P.}~\bibnamefont
			{Kirton}}\ and\ \bibinfo {author} {\bibfnamefont {J.}~\bibnamefont
			{Keeling}},\ }\bibfield  {title} {\bibinfo {title} {Superradiant and lasing
			states in driven-dissipative {Dicke} models},\ }\href
	{https://doi.org/10.1088/1367-2630/aaa11d} {\bibfield  {journal} {\bibinfo
			{journal} {New J. Phys.}\ }\textbf {\bibinfo {volume} {20}},\ \bibinfo
		{pages} {015009} (\bibinfo {year} {2018})}\BibitemShut {NoStop}%
	\bibitem [{\citenamefont {Shammah}\ \emph {et~al.}(2018)\citenamefont
		{Shammah}, \citenamefont {Ahmed}, \citenamefont {Lambert}, \citenamefont
		{De~Liberato},\ and\ \citenamefont {Nori}}]{PhysRevA.98.063815}%
	\BibitemOpen
	\bibfield  {author} {\bibinfo {author} {\bibfnamefont {N.}~\bibnamefont
			{Shammah}}, \bibinfo {author} {\bibfnamefont {S.}~\bibnamefont {Ahmed}},
		\bibinfo {author} {\bibfnamefont {N.}~\bibnamefont {Lambert}}, \bibinfo
		{author} {\bibfnamefont {S.}~\bibnamefont {De~Liberato}},\ and\ \bibinfo
		{author} {\bibfnamefont {F.}~\bibnamefont {Nori}},\ }\bibfield  {title}
	{\bibinfo {title} {Open quantum systems with local and collective incoherent
			processes: Efficient numerical simulations using permutational invariance},\
	}\href {https://doi.org/10.1103/PhysRevA.98.063815} {\bibfield  {journal}
		{\bibinfo  {journal} {Phys. Rev. A}\ }\textbf {\bibinfo {volume} {98}},\
		\bibinfo {pages} {063815} (\bibinfo {year} {2018})}\BibitemShut {NoStop}%
	\bibitem [{\citenamefont {Damanet}\ \emph {et~al.}(2019)\citenamefont
		{Damanet}, \citenamefont {Daley},\ and\ \citenamefont
		{Keeling}}]{PhysRevA.99.033845}%
	\BibitemOpen
	\bibfield  {author} {\bibinfo {author} {\bibfnamefont {F.}~\bibnamefont
			{Damanet}}, \bibinfo {author} {\bibfnamefont {A.~J.}\ \bibnamefont {Daley}},\
		and\ \bibinfo {author} {\bibfnamefont {J.}~\bibnamefont {Keeling}},\
	}\bibfield  {title} {\bibinfo {title} {Atom-only descriptions of the
			driven-dissipative {Dicke} model},\ }\href
	{https://doi.org/10.1103/PhysRevA.99.033845} {\bibfield  {journal} {\bibinfo
			{journal} {Phys. Rev. A}\ }\textbf {\bibinfo {volume} {99}},\ \bibinfo
		{pages} {033845} (\bibinfo {year} {2019})}\BibitemShut {NoStop}%
	\bibitem [{\citenamefont {Bonifacio}\ \emph
		{et~al.}(1971{\natexlab{a}})\citenamefont {Bonifacio}, \citenamefont
		{Schwendimann},\ and\ \citenamefont {Haake}}]{Bonifacio1971(I)}%
	\BibitemOpen
	\bibfield  {author} {\bibinfo {author} {\bibfnamefont {R.}~\bibnamefont
			{Bonifacio}}, \bibinfo {author} {\bibfnamefont {P.}~\bibnamefont
			{Schwendimann}},\ and\ \bibinfo {author} {\bibfnamefont {F.}~\bibnamefont
			{Haake}},\ }\bibfield  {title} {\bibinfo {title} {Quantum {statistical}
			{theory} of {superradiance}. {I}},\ }\href
	{https://doi.org/10.1103/PhysRevA.4.302} {\bibfield  {journal} {\bibinfo
			{journal} {Phys. Rev. A}\ }\textbf {\bibinfo {volume} {4}},\ \bibinfo {pages}
		{302} (\bibinfo {year} {1971}{\natexlab{a}})}\BibitemShut {NoStop}%
	\bibitem [{\citenamefont {Bonifacio}\ \emph
		{et~al.}(1971{\natexlab{b}})\citenamefont {Bonifacio}, \citenamefont
		{Schwendimann},\ and\ \citenamefont {Haake}}]{Bonifacio1971(II)}%
	\BibitemOpen
	\bibfield  {author} {\bibinfo {author} {\bibfnamefont {R.}~\bibnamefont
			{Bonifacio}}, \bibinfo {author} {\bibfnamefont {P.}~\bibnamefont
			{Schwendimann}},\ and\ \bibinfo {author} {\bibfnamefont {F.}~\bibnamefont
			{Haake}},\ }\bibfield  {title} {\bibinfo {title} {Quantum {statistical}
			{theory} of {superradiance}. {II}},\ }\href
	{https://doi.org/10.1103/PhysRevA.4.854} {\bibfield  {journal} {\bibinfo
			{journal} {Phys. Rev. A}\ }\textbf {\bibinfo {volume} {4}},\ \bibinfo {pages}
		{854} (\bibinfo {year} {1971}{\natexlab{b}})}\BibitemShut {NoStop}%
	\bibitem [{\citenamefont {Azouit}\ \emph {et~al.}(2017)\citenamefont {Azouit},
		\citenamefont {Chittaro}, \citenamefont {Sarlette},\ and\ \citenamefont
		{Rouchon}}]{Azouit2017}%
	\BibitemOpen
	\bibfield  {author} {\bibinfo {author} {\bibfnamefont {R.}~\bibnamefont
			{Azouit}}, \bibinfo {author} {\bibfnamefont {F.}~\bibnamefont {Chittaro}},
		\bibinfo {author} {\bibfnamefont {A.}~\bibnamefont {Sarlette}},\ and\
		\bibinfo {author} {\bibfnamefont {P.}~\bibnamefont {Rouchon}},\ }\bibfield
	{title} {\bibinfo {title} {Towards generic adiabatic elimination for
			bipartite open quantum systems},\ }\href
	{https://doi.org/10.1088/2058-9565/aa7f3f} {\bibfield  {journal} {\bibinfo
			{journal} {Quantum Science and Technology}\ }\textbf {\bibinfo {volume}
			{2}},\ \bibinfo {pages} {044011} (\bibinfo {year} {2017})}\BibitemShut
	{NoStop}%
	\bibitem [{\citenamefont {Azouit}(2017)}]{ThesisAzouit2017}%
	\BibitemOpen
	\bibfield  {author} {\bibinfo {author} {\bibfnamefont {R.}~\bibnamefont
			{Azouit}},\ }\emph {\bibinfo {title} {{Adiabatic elimination for open quantum
				systems}}},\ \href {https://pastel.archives-ouvertes.fr/tel-01743808}
	{\bibinfo {type} {Theses}},\ \bibinfo  {school} {{PSL Research University}}
	(\bibinfo {year} {2017})\BibitemShut {NoStop}%
	\bibitem [{\citenamefont {Leroux}\ \emph {et~al.}(2010)\citenamefont {Leroux},
		\citenamefont {Schleier-Smith},\ and\ \citenamefont
		{Vuleti\ifmmode~\acute{c}\else \'{c}\fi{}}}]{PhysRevLett.104.073602}%
	\BibitemOpen
	\bibfield  {author} {\bibinfo {author} {\bibfnamefont {I.~D.}\ \bibnamefont
			{Leroux}}, \bibinfo {author} {\bibfnamefont {M.~H.}\ \bibnamefont
			{Schleier-Smith}},\ and\ \bibinfo {author} {\bibfnamefont {V.}~\bibnamefont
			{Vuleti\ifmmode~\acute{c}\else \'{c}\fi{}}},\ }\bibfield  {title} {\bibinfo
		{title} {Implementation of {Cavity} {Squeezing} of a {Collective} {Atomic}
			{Spin}},\ }\href {https://doi.org/10.1103/PhysRevLett.104.073602} {\bibfield
		{journal} {\bibinfo  {journal} {Phys. Rev. Lett.}\ }\textbf {\bibinfo
			{volume} {104}},\ \bibinfo {pages} {073602} (\bibinfo {year}
		{2010})}\BibitemShut {NoStop}%
	\bibitem [{\citenamefont {T{\'o}th}\ and\ \citenamefont
		{Apellaniz}(2014)}]{toth2014quantum}%
	\BibitemOpen
	\bibfield  {author} {\bibinfo {author} {\bibfnamefont {G.}~\bibnamefont
			{T{\'o}th}}\ and\ \bibinfo {author} {\bibfnamefont {I.}~\bibnamefont
			{Apellaniz}},\ }\bibfield  {title} {\bibinfo {title} {Quantum metrology from
			a quantum information science perspective},\ }\href
	{https://doi.org/10.1088/1751-8113/47/42/424006} {\bibfield  {journal}
		{\bibinfo  {journal} {J. Phys. A}\ }\textbf {\bibinfo {volume} {47}},\
		\bibinfo {pages} {424006} (\bibinfo {year} {2014})}\BibitemShut {NoStop}%
	\bibitem [{\citenamefont {Jahnke}\ \emph {et~al.}(2016)\citenamefont {Jahnke},
		\citenamefont {Gies}, \citenamefont {A{\ss}mann}, \citenamefont {Bayer},
		\citenamefont {Leymann}, \citenamefont {Foerster}, \citenamefont {Wiersig},
		\citenamefont {Schneider}, \citenamefont {Kamp},\ and\ \citenamefont
		{H{\"o}fling}}]{jahnke2016giant}%
	\BibitemOpen
	\bibfield  {author} {\bibinfo {author} {\bibfnamefont {F.}~\bibnamefont
			{Jahnke}}, \bibinfo {author} {\bibfnamefont {C.}~\bibnamefont {Gies}},
		\bibinfo {author} {\bibfnamefont {M.}~\bibnamefont {A{\ss}mann}}, \bibinfo
		{author} {\bibfnamefont {M.}~\bibnamefont {Bayer}}, \bibinfo {author}
		{\bibfnamefont {H.}~\bibnamefont {Leymann}}, \bibinfo {author} {\bibfnamefont
			{A.}~\bibnamefont {Foerster}}, \bibinfo {author} {\bibfnamefont
			{J.}~\bibnamefont {Wiersig}}, \bibinfo {author} {\bibfnamefont
			{C.}~\bibnamefont {Schneider}}, \bibinfo {author} {\bibfnamefont
			{M.}~\bibnamefont {Kamp}},\ and\ \bibinfo {author} {\bibfnamefont
			{S.}~\bibnamefont {H{\"o}fling}},\ }\bibfield  {title} {\bibinfo {title}
		{Giant photon bunching, superradiant pulse emission and excitation trapping
			in quantum-dot nanolasers},\ }\href
	{https://doi.org/https://doi.org/10.1038/ncomms11540} {\bibfield  {journal}
		{\bibinfo  {journal} {Nat. Commun.}\ }\textbf {\bibinfo {volume} {7}},\
		\bibinfo {pages} {11540} (\bibinfo {year} {2016})}\BibitemShut {NoStop}%
	\bibitem [{\citenamefont {Asenjo-Garcia}\ \emph {et~al.}(2017)\citenamefont
		{Asenjo-Garcia}, \citenamefont {Moreno-Cardoner}, \citenamefont {Albrecht},
		\citenamefont {Kimble},\ and\ \citenamefont {Chang}}]{PhysRevX.7.031024}%
	\BibitemOpen
	\bibfield  {author} {\bibinfo {author} {\bibfnamefont {A.}~\bibnamefont
			{Asenjo-Garcia}}, \bibinfo {author} {\bibfnamefont {M.}~\bibnamefont
			{Moreno-Cardoner}}, \bibinfo {author} {\bibfnamefont {A.}~\bibnamefont
			{Albrecht}}, \bibinfo {author} {\bibfnamefont {H.~J.}\ \bibnamefont
			{Kimble}},\ and\ \bibinfo {author} {\bibfnamefont {D.~E.}\ \bibnamefont
			{Chang}},\ }\bibfield  {title} {\bibinfo {title} {Exponential {Improvement}
			in {Photon} {Storage} {Fidelities} {Using} {Subradiance} and {``Selective
				Radiance''} in {Atomic} {Arrays}},\ }\href
	{https://doi.org/10.1103/PhysRevX.7.031024} {\bibfield  {journal} {\bibinfo
			{journal} {Phys. Rev. X}\ }\textbf {\bibinfo {volume} {7}},\ \bibinfo {pages}
		{031024} (\bibinfo {year} {2017})}\BibitemShut {NoStop}%
	\bibitem [{\citenamefont {Garcia-Vidal}\ and\ \citenamefont
		{Feist}(2017)}]{garcia2017long}%
	\BibitemOpen
	\bibfield  {author} {\bibinfo {author} {\bibfnamefont {F.~J.}\ \bibnamefont
			{Garcia-Vidal}}\ and\ \bibinfo {author} {\bibfnamefont {J.}~\bibnamefont
			{Feist}},\ }\bibfield  {title} {\bibinfo {title} {Long-distance operator for
			energy transfer},\ }\href {https://doi.org/10.1126/science.aao4268}
	{\bibfield  {journal} {\bibinfo  {journal} {Science}\ }\textbf {\bibinfo
			{volume} {357}},\ \bibinfo {pages} {1357} (\bibinfo {year}
		{2017})}\BibitemShut {NoStop}%
	\bibitem [{\citenamefont {Goban}\ \emph {et~al.}(2015)\citenamefont {Goban},
		\citenamefont {Hung}, \citenamefont {Hood}, \citenamefont {Yu}, \citenamefont
		{Muniz}, \citenamefont {Painter},\ and\ \citenamefont
		{Kimble}}]{PhysRevLett.115.063601}%
	\BibitemOpen
	\bibfield  {author} {\bibinfo {author} {\bibfnamefont {A.}~\bibnamefont
			{Goban}}, \bibinfo {author} {\bibfnamefont {C.-L.}\ \bibnamefont {Hung}},
		\bibinfo {author} {\bibfnamefont {J.~D.}\ \bibnamefont {Hood}}, \bibinfo
		{author} {\bibfnamefont {S.-P.}\ \bibnamefont {Yu}}, \bibinfo {author}
		{\bibfnamefont {J.~A.}\ \bibnamefont {Muniz}}, \bibinfo {author}
		{\bibfnamefont {O.}~\bibnamefont {Painter}},\ and\ \bibinfo {author}
		{\bibfnamefont {H.~J.}\ \bibnamefont {Kimble}},\ }\bibfield  {title}
	{\bibinfo {title} {Superradiance for {Atoms} {Trapped} along a {Photonic}
			{Crystal} {Waveguide}},\ }\href
	{https://doi.org/10.1103/PhysRevLett.115.063601} {\bibfield  {journal}
		{\bibinfo  {journal} {Phys. Rev. Lett.}\ }\textbf {\bibinfo {volume} {115}},\
		\bibinfo {pages} {063601} (\bibinfo {year} {2015})}\BibitemShut {NoStop}%
	\bibitem [{\citenamefont {DeVoe}\ and\ \citenamefont
		{Brewer}(1996)}]{PhysRevLett.76.2049}%
	\BibitemOpen
	\bibfield  {author} {\bibinfo {author} {\bibfnamefont {R.~G.}\ \bibnamefont
			{DeVoe}}\ and\ \bibinfo {author} {\bibfnamefont {R.~G.}\ \bibnamefont
			{Brewer}},\ }\bibfield  {title} {\bibinfo {title} {Observation of
			{Superradiant} and {Subradiant} {Spontaneous} {Emission} of {Two} {Trapped}
			{Ions}},\ }\href {https://doi.org/10.1103/PhysRevLett.76.2049} {\bibfield
		{journal} {\bibinfo  {journal} {Phys. Rev. Lett.}\ }\textbf {\bibinfo
			{volume} {76}},\ \bibinfo {pages} {2049} (\bibinfo {year}
		{1996})}\BibitemShut {NoStop}%
	\bibitem [{\citenamefont {Jenkins}\ \emph {et~al.}(2017)\citenamefont
		{Jenkins}, \citenamefont {Ruostekoski}, \citenamefont {Papasimakis},
		\citenamefont {Savo},\ and\ \citenamefont
		{Zheludev}}]{PhysRevLett.119.053901}%
	\BibitemOpen
	\bibfield  {author} {\bibinfo {author} {\bibfnamefont {S.~D.}\ \bibnamefont
			{Jenkins}}, \bibinfo {author} {\bibfnamefont {J.}~\bibnamefont
			{Ruostekoski}}, \bibinfo {author} {\bibfnamefont {N.}~\bibnamefont
			{Papasimakis}}, \bibinfo {author} {\bibfnamefont {S.}~\bibnamefont {Savo}},\
		and\ \bibinfo {author} {\bibfnamefont {N.~I.}\ \bibnamefont {Zheludev}},\
	}\bibfield  {title} {\bibinfo {title} {{Many-Body} {Subradiant} {Excitations}
			in {Metamaterial} {Arrays}: {Experiment} and {Theory}},\ }\href
	{https://doi.org/10.1103/PhysRevLett.119.053901} {\bibfield  {journal}
		{\bibinfo  {journal} {Phys. Rev. Lett.}\ }\textbf {\bibinfo {volume} {119}},\
		\bibinfo {pages} {053901} (\bibinfo {year} {2017})}\BibitemShut {NoStop}%
	\bibitem [{\citenamefont {Pustovit}\ and\ \citenamefont
		{Shahbazyan}(2009)}]{PhysRevLett.102.077401}%
	\BibitemOpen
	\bibfield  {author} {\bibinfo {author} {\bibfnamefont {V.~N.}\ \bibnamefont
			{Pustovit}}\ and\ \bibinfo {author} {\bibfnamefont {T.~V.}\ \bibnamefont
			{Shahbazyan}},\ }\bibfield  {title} {\bibinfo {title} {Cooperative Emission
			of Light by an Ensemble of Dipoles Near a Metal Nanoparticle: The Plasmonic
			{Dicke} Effect},\ }\href {https://doi.org/10.1103/PhysRevLett.102.077401}
	{\bibfield  {journal} {\bibinfo  {journal} {Phys. Rev. Lett.}\ }\textbf
		{\bibinfo {volume} {102}},\ \bibinfo {pages} {077401} (\bibinfo {year}
		{2009})}\BibitemShut {NoStop}%
	\bibitem [{\citenamefont {Angerer}\ \emph {et~al.}(2018)\citenamefont {Angerer}
		\emph {et~al.}}]{angerer2018superradiant}%
	\BibitemOpen
	\bibfield  {author} {\bibinfo {author} {\bibfnamefont {A.}~\bibnamefont
			{Angerer}} \emph {et~al.},\ }\bibfield  {title} {\bibinfo {title}
		{Superradiant emission from colour centres in diamond},\ }\href
	{https://doi.org/https://doi.org/10.1038/s41567-018-0269-7} {\bibfield
		{journal} {\bibinfo  {journal} {Nat. Phys.}\ }\textbf {\bibinfo {volume}
			{14}},\ \bibinfo {pages} {1168} (\bibinfo {year} {2018})}\BibitemShut
	{NoStop}%
	\bibitem [{\citenamefont {Scheibner}\ \emph {et~al.}(2007)\citenamefont
		{Scheibner}, \citenamefont {Schmidt}, \citenamefont {Worschech},
		\citenamefont {Forchel}, \citenamefont {Bacher}, \citenamefont {Passow},\
		and\ \citenamefont {Hommel}}]{scheibner2007superradiance}%
	\BibitemOpen
	\bibfield  {author} {\bibinfo {author} {\bibfnamefont {M.}~\bibnamefont
			{Scheibner}}, \bibinfo {author} {\bibfnamefont {T.}~\bibnamefont {Schmidt}},
		\bibinfo {author} {\bibfnamefont {L.}~\bibnamefont {Worschech}}, \bibinfo
		{author} {\bibfnamefont {A.}~\bibnamefont {Forchel}}, \bibinfo {author}
		{\bibfnamefont {G.}~\bibnamefont {Bacher}}, \bibinfo {author} {\bibfnamefont
			{T.}~\bibnamefont {Passow}},\ and\ \bibinfo {author} {\bibfnamefont
			{D.}~\bibnamefont {Hommel}},\ }\bibfield  {title} {\bibinfo {title}
		{Superradiance of quantum dots},\ }\href
	{https://doi.org/https://doi.org/10.1038/nphys494} {\bibfield  {journal}
		{\bibinfo  {journal} {Nat. Phys.}\ }\textbf {\bibinfo {volume} {3}},\
		\bibinfo {pages} {106} (\bibinfo {year} {2007})}\BibitemShut {NoStop}%
	\bibitem [{\citenamefont {Mlynek}\ \emph {et~al.}(2014)\citenamefont {Mlynek},
		\citenamefont {Abdumalikov}, \citenamefont {Eichler},\ and\ \citenamefont
		{Wallraff}}]{mlynek2014observation}%
	\BibitemOpen
	\bibfield  {author} {\bibinfo {author} {\bibfnamefont {J.~A.}\ \bibnamefont
			{Mlynek}}, \bibinfo {author} {\bibfnamefont {A.~A.}\ \bibnamefont
			{Abdumalikov}}, \bibinfo {author} {\bibfnamefont {C.}~\bibnamefont
			{Eichler}},\ and\ \bibinfo {author} {\bibfnamefont {A.}~\bibnamefont
			{Wallraff}},\ }\bibfield  {title} {\bibinfo {title} {Observation of {Dicke}
			superradiance for two artificial atoms in a cavity with high decay rate},\
	}\href {https://doi.org/https://doi.org/10.1038/ncomms6186} {\bibfield
		{journal} {\bibinfo  {journal} {Nat. Commun.}\ }\textbf {\bibinfo {volume}
			{5}},\ \bibinfo {pages} {5186} (\bibinfo {year} {2014})}\BibitemShut
	{NoStop}%
	\bibitem [{\citenamefont {Felicetti}\ \emph
		{et~al.}(2018{\natexlab{a}})\citenamefont {Felicetti}, \citenamefont
		{Rossatto}, \citenamefont {Rico}, \citenamefont {Solano},\ and\ \citenamefont
		{Forn-D\'{\i}az}}]{felicetti_two-photon_2018}%
	\BibitemOpen
	\bibfield  {author} {\bibinfo {author} {\bibfnamefont {S.}~\bibnamefont
			{Felicetti}}, \bibinfo {author} {\bibfnamefont {D.~Z.}\ \bibnamefont
			{Rossatto}}, \bibinfo {author} {\bibfnamefont {E.}~\bibnamefont {Rico}},
		\bibinfo {author} {\bibfnamefont {E.}~\bibnamefont {Solano}},\ and\ \bibinfo
		{author} {\bibfnamefont {P.}~\bibnamefont {Forn-D\'{\i}az}},\ }\bibfield
	{title} {\bibinfo {title} {Two-photon quantum {Rabi} model with
			superconducting circuits},\ }\href
	{https://doi.org/10.1103/PhysRevA.97.013851} {\bibfield  {journal} {\bibinfo
			{journal} {Phys. Rev. A}\ }\textbf {\bibinfo {volume} {97}},\ \bibinfo
		{pages} {013851} (\bibinfo {year} {2018}{\natexlab{a}})}\BibitemShut
	{NoStop}%
	\bibitem [{\citenamefont {Felicetti}\ \emph
		{et~al.}(2018{\natexlab{b}})\citenamefont {Felicetti}, \citenamefont
		{Hwang},\ and\ \citenamefont {Le~Boit\'e}}]{PhysRevA.98.053859}%
	\BibitemOpen
	\bibfield  {author} {\bibinfo {author} {\bibfnamefont {S.}~\bibnamefont
			{Felicetti}}, \bibinfo {author} {\bibfnamefont {M.-J.}\ \bibnamefont
			{Hwang}},\ and\ \bibinfo {author} {\bibfnamefont {A.}~\bibnamefont
			{Le~Boit\'e}},\ }\bibfield  {title} {\bibinfo {title} {Ultrastrong-coupling
			regime of nondipolar light-matter interactions},\ }\href
	{https://doi.org/10.1103/PhysRevA.98.053859} {\bibfield  {journal} {\bibinfo
			{journal} {Phys. Rev. A}\ }\textbf {\bibinfo {volume} {98}},\ \bibinfo
		{pages} {053859} (\bibinfo {year} {2018}{\natexlab{b}})}\BibitemShut
	{NoStop}%
	\bibitem [{\citenamefont {Felicetti}\ \emph {et~al.}(2015)\citenamefont
		{Felicetti}, \citenamefont {Pedernales}, \citenamefont {Egusquiza},
		\citenamefont {Romero}, \citenamefont {Lamata}, \citenamefont {Braak},\ and\
		\citenamefont {Solano}}]{felicetti_spectral_2015}%
	\BibitemOpen
	\bibfield  {author} {\bibinfo {author} {\bibfnamefont {S.}~\bibnamefont
			{Felicetti}}, \bibinfo {author} {\bibfnamefont {J.~S.}\ \bibnamefont
			{Pedernales}}, \bibinfo {author} {\bibfnamefont {I.~L.}\ \bibnamefont
			{Egusquiza}}, \bibinfo {author} {\bibfnamefont {G.}~\bibnamefont {Romero}},
		\bibinfo {author} {\bibfnamefont {L.}~\bibnamefont {Lamata}}, \bibinfo
		{author} {\bibfnamefont {D.}~\bibnamefont {Braak}},\ and\ \bibinfo {author}
		{\bibfnamefont {E.}~\bibnamefont {Solano}},\ }\bibfield  {title} {\bibinfo
		{title} {Spectral collapse via two-phonon interactions in trapped ions},\
	}\href {https://doi.org/10.1103/PhysRevA.92.033817} {\bibfield  {journal}
		{\bibinfo  {journal} {Phys. Rev. A}\ }\textbf {\bibinfo {volume} {92}},\
		\bibinfo {pages} {033817} (\bibinfo {year} {2015})}\BibitemShut {NoStop}%
	\bibitem [{\citenamefont {Cheng}\ \emph {et~al.}(2018)\citenamefont {Cheng},
		\citenamefont {Arrazola}, \citenamefont {Pedernales}, \citenamefont {Lamata},
		\citenamefont {Chen},\ and\ \citenamefont {Solano}}]{PhysRevA.97.023624}%
	\BibitemOpen
	\bibfield  {author} {\bibinfo {author} {\bibfnamefont {X.-H.}\ \bibnamefont
			{Cheng}}, \bibinfo {author} {\bibfnamefont {I.}~\bibnamefont {Arrazola}},
		\bibinfo {author} {\bibfnamefont {J.~S.}\ \bibnamefont {Pedernales}},
		\bibinfo {author} {\bibfnamefont {L.}~\bibnamefont {Lamata}}, \bibinfo
		{author} {\bibfnamefont {X.}~\bibnamefont {Chen}},\ and\ \bibinfo {author}
		{\bibfnamefont {E.}~\bibnamefont {Solano}},\ }\bibfield  {title} {\bibinfo
		{title} {Nonlinear quantum {Rabi} model in trapped ions},\ }\href
	{https://doi.org/10.1103/PhysRevA.97.023624} {\bibfield  {journal} {\bibinfo
			{journal} {Phys. Rev. A}\ }\textbf {\bibinfo {volume} {97}},\ \bibinfo
		{pages} {023624} (\bibinfo {year} {2018})}\BibitemShut {NoStop}%
	\bibitem [{\citenamefont {Puebla}\ \emph {et~al.}(2019)\citenamefont {Puebla},
		\citenamefont {Casanova}, \citenamefont {Houhou}, \citenamefont {Solano},\
		and\ \citenamefont {Paternostro}}]{PhysRevA.99.032303}%
	\BibitemOpen
	\bibfield  {author} {\bibinfo {author} {\bibfnamefont {R.}~\bibnamefont
			{Puebla}}, \bibinfo {author} {\bibfnamefont {J.}~\bibnamefont {Casanova}},
		\bibinfo {author} {\bibfnamefont {O.}~\bibnamefont {Houhou}}, \bibinfo
		{author} {\bibfnamefont {E.}~\bibnamefont {Solano}},\ and\ \bibinfo {author}
		{\bibfnamefont {M.}~\bibnamefont {Paternostro}},\ }\bibfield  {title}
	{\bibinfo {title} {Quantum simulation of multiphoton and nonlinear
			dissipative spin-boson models},\ }\href
	{https://doi.org/10.1103/PhysRevA.99.032303} {\bibfield  {journal} {\bibinfo
			{journal} {Phys. Rev. A}\ }\textbf {\bibinfo {volume} {99}},\ \bibinfo
		{pages} {032303} (\bibinfo {year} {2019})}\BibitemShut {NoStop}%
	\bibitem [{\citenamefont {Schneeweiss}\ \emph {et~al.}(2018)\citenamefont
		{Schneeweiss}, \citenamefont {Dareau},\ and\ \citenamefont
		{Sayrin}}]{Schneeweiss_2018}%
	\BibitemOpen
	\bibfield  {author} {\bibinfo {author} {\bibfnamefont {P.}~\bibnamefont
			{Schneeweiss}}, \bibinfo {author} {\bibfnamefont {A.}~\bibnamefont
			{Dareau}},\ and\ \bibinfo {author} {\bibfnamefont {C.}~\bibnamefont
			{Sayrin}},\ }\bibfield  {title} {\bibinfo {title} {Cold-atom-based
			implementation of the quantum {Rabi} model},\ }\href
	{https://doi.org/10.1103/PhysRevA.98.021801} {\bibfield  {journal} {\bibinfo
			{journal} {Phys. Rev. A}\ }\textbf {\bibinfo {volume} {98}},\ \bibinfo
		{pages} {021801(R)} (\bibinfo {year} {2018})}\BibitemShut {NoStop}%
	\bibitem [{\citenamefont {Dareau}\ \emph {et~al.}(2018)\citenamefont {Dareau},
		\citenamefont {Meng}, \citenamefont {Schneeweiss},\ and\ \citenamefont
		{Rauschenbeutel}}]{Dareau_2018}%
	\BibitemOpen
	\bibfield  {author} {\bibinfo {author} {\bibfnamefont {A.}~\bibnamefont
			{Dareau}}, \bibinfo {author} {\bibfnamefont {Y.}~\bibnamefont {Meng}},
		\bibinfo {author} {\bibfnamefont {P.}~\bibnamefont {Schneeweiss}},\ and\
		\bibinfo {author} {\bibfnamefont {A.}~\bibnamefont {Rauschenbeutel}},\
	}\bibfield  {title} {\bibinfo {title} {Observation of {Ultrastrong}
			{Spin-Motion} {Coupling} for {Cold} {Atoms} in {Optical} {Microtraps}},\
	}\href {https://doi.org/10.1103/PhysRevLett.121.253603} {\bibfield  {journal}
		{\bibinfo  {journal} {Phys. Rev. Lett.}\ }\textbf {\bibinfo {volume} {121}},\
		\bibinfo {pages} {253603} (\bibinfo {year} {2018})}\BibitemShut {NoStop}%
	\bibitem [{\citenamefont {Goetz}\ \emph {et~al.}(2018)\citenamefont {Goetz},
		\citenamefont {Deppe}, \citenamefont {Fedorov}, \citenamefont {Eder},
		\citenamefont {Fischer}, \citenamefont {Pogorzalek}, \citenamefont {Xie},
		\citenamefont {Marx},\ and\ \citenamefont {Gross}}]{PhysRevLett.121.060503}%
	\BibitemOpen
	\bibfield  {author} {\bibinfo {author} {\bibfnamefont {J.}~\bibnamefont
			{Goetz}}, \bibinfo {author} {\bibfnamefont {F.}~\bibnamefont {Deppe}},
		\bibinfo {author} {\bibfnamefont {K.~G.}\ \bibnamefont {Fedorov}}, \bibinfo
		{author} {\bibfnamefont {P.}~\bibnamefont {Eder}}, \bibinfo {author}
		{\bibfnamefont {M.}~\bibnamefont {Fischer}}, \bibinfo {author} {\bibfnamefont
			{S.}~\bibnamefont {Pogorzalek}}, \bibinfo {author} {\bibfnamefont
			{E.}~\bibnamefont {Xie}}, \bibinfo {author} {\bibfnamefont {A.}~\bibnamefont
			{Marx}},\ and\ \bibinfo {author} {\bibfnamefont {R.}~\bibnamefont {Gross}},\
	}\bibfield  {title} {\bibinfo {title} {{Parity-Engineered} {Light-Matter}
			{Interaction}},\ }\href {https://doi.org/10.1103/PhysRevLett.121.060503}
	{\bibfield  {journal} {\bibinfo  {journal} {Phys. Rev. Lett.}\ }\textbf
		{\bibinfo {volume} {121}},\ \bibinfo {pages} {060503} (\bibinfo {year}
		{2018})}\BibitemShut {NoStop}%
	\bibitem [{\citenamefont {Lv}\ \emph {et~al.}(2018)\citenamefont {Lv},
		\citenamefont {An}, \citenamefont {Liu}, \citenamefont {Zhang}, \citenamefont
		{Pedernales}, \citenamefont {Lamata}, \citenamefont {Solano},\ and\
		\citenamefont {Kim}}]{PhysRevX.8.021027}%
	\BibitemOpen
	\bibfield  {author} {\bibinfo {author} {\bibfnamefont {D.}~\bibnamefont
			{Lv}}, \bibinfo {author} {\bibfnamefont {S.}~\bibnamefont {An}}, \bibinfo
		{author} {\bibfnamefont {Z.}~\bibnamefont {Liu}}, \bibinfo {author}
		{\bibfnamefont {J.-N.}\ \bibnamefont {Zhang}}, \bibinfo {author}
		{\bibfnamefont {J.~S.}\ \bibnamefont {Pedernales}}, \bibinfo {author}
		{\bibfnamefont {L.}~\bibnamefont {Lamata}}, \bibinfo {author} {\bibfnamefont
			{E.}~\bibnamefont {Solano}},\ and\ \bibinfo {author} {\bibfnamefont
			{K.}~\bibnamefont {Kim}},\ }\bibfield  {title} {\bibinfo {title} {Quantum
			{Simulation} of the {Quantum} {Rabi} {Model} in a {Trapped} {Ion}},\ }\href
	{https://doi.org/10.1103/PhysRevX.8.021027} {\bibfield  {journal} {\bibinfo
			{journal} {Phys. Rev. X}\ }\textbf {\bibinfo {volume} {8}},\ \bibinfo {pages}
		{021027} (\bibinfo {year} {2018})}\BibitemShut {NoStop}%
	\bibitem [{\citenamefont {Minganti}\ \emph {et~al.}(2016)\citenamefont
		{Minganti}, \citenamefont {Bartolo}, \citenamefont {Lolli}, \citenamefont
		{Casteels},\ and\ \citenamefont {Ciuti}}]{Minganti2016}%
	\BibitemOpen
	\bibfield  {author} {\bibinfo {author} {\bibfnamefont {F.}~\bibnamefont
			{Minganti}}, \bibinfo {author} {\bibfnamefont {N.}~\bibnamefont {Bartolo}},
		\bibinfo {author} {\bibfnamefont {J.}~\bibnamefont {Lolli}}, \bibinfo
		{author} {\bibfnamefont {W.}~\bibnamefont {Casteels}},\ and\ \bibinfo
		{author} {\bibfnamefont {C.}~\bibnamefont {Ciuti}},\ }\bibfield  {title}
	{\bibinfo {title} {Exact results for {Schr{\"o}dinger} cats in
			driven-dissipative systems and their feedback control},\ }\href
	{https://www.nature.com/articles/srep26987} {\bibfield  {journal} {\bibinfo
			{journal} {Scientific reports}\ }\textbf {\bibinfo {volume} {6}},\ \bibinfo
		{pages} {26987} (\bibinfo {year} {2016})}\BibitemShut {NoStop}%
	\bibitem [{\citenamefont {Malekakhlagh}\ and\ \citenamefont
		{Rodriguez}(2019)}]{Malek2019}%
	\BibitemOpen
	\bibfield  {author} {\bibinfo {author} {\bibfnamefont {M.}~\bibnamefont
			{Malekakhlagh}}\ and\ \bibinfo {author} {\bibfnamefont {A.~W.}\ \bibnamefont
			{Rodriguez}},\ }\bibfield  {title} {\bibinfo {title} {{Quantum} {Rabi}
			{Model} with {Two-Photon} {Relaxation}},\ }\href
	{https://doi.org/10.1103/PhysRevLett.122.043601} {\bibfield  {journal}
		{\bibinfo  {journal} {Phys. Rev. Lett.}\ }\textbf {\bibinfo {volume} {122}},\
		\bibinfo {pages} {043601} (\bibinfo {year} {2019})}\BibitemShut {NoStop}%
	\bibitem [{\citenamefont {Leghtas}\ \emph {et~al.}(2015)\citenamefont {Leghtas}
		\emph {et~al.}}]{Leghtas2015}%
	\BibitemOpen
	\bibfield  {author} {\bibinfo {author} {\bibfnamefont {Z.}~\bibnamefont
			{Leghtas}} \emph {et~al.},\ }\bibfield  {title} {\bibinfo {title} {Confining
			the state of light to a quantum manifold by engineered two-photon loss},\
	}\href {https://doi.org/10.1126/science.aaa2085} {\bibfield  {journal}
		{\bibinfo  {journal} {Science}\ }\textbf {\bibinfo {volume} {347}},\ \bibinfo
		{pages} {853} (\bibinfo {year} {2015})}\BibitemShut {NoStop}%
	\bibitem [{\citenamefont {Trav\ifmmode~\check{e}\else
			\v{e}\fi{}nec}(2012)}]{PhysRevA.85.043805}%
	\BibitemOpen
	\bibfield  {author} {\bibinfo {author} {\bibfnamefont {I.}~\bibnamefont
			{Trav\ifmmode~\check{e}\else \v{e}\fi{}nec}},\ }\bibfield  {title} {\bibinfo
		{title} {Solvability of the two-photon {Rabi} {Hamiltonian}},\ }\href
	{https://doi.org/10.1103/PhysRevA.85.043805} {\bibfield  {journal} {\bibinfo
			{journal} {Phys. Rev. A}\ }\textbf {\bibinfo {volume} {85}},\ \bibinfo
		{pages} {043805} (\bibinfo {year} {2012})}\BibitemShut {NoStop}%
	\bibitem [{\citenamefont {Duan}\ \emph {et~al.}(2016)\citenamefont {Duan},
		\citenamefont {Xie}, \citenamefont {Braak},\ and\ \citenamefont
		{Chen}}]{duan2016two}%
	\BibitemOpen
	\bibfield  {author} {\bibinfo {author} {\bibfnamefont {L.}~\bibnamefont
			{Duan}}, \bibinfo {author} {\bibfnamefont {Y.-F.}\ \bibnamefont {Xie}},
		\bibinfo {author} {\bibfnamefont {D.}~\bibnamefont {Braak}},\ and\ \bibinfo
		{author} {\bibfnamefont {Q.-H.}\ \bibnamefont {Chen}},\ }\bibfield  {title}
	{\bibinfo {title} {Two-photon {Rabi} model: Analytic solutions and spectral
			collapse},\ }\href {https://doi.org/10.1088/1751-8113/49/46/464002}
	{\bibfield  {journal} {\bibinfo  {journal} {J. Phys. A}\ }\textbf {\bibinfo
			{volume} {49}},\ \bibinfo {pages} {464002} (\bibinfo {year}
		{2016})}\BibitemShut {NoStop}%
	\bibitem [{\citenamefont {Maciejewski}\ and\ \citenamefont
		{Stachowiak}(2017)}]{maciejewski2017novel}%
	\BibitemOpen
	\bibfield  {author} {\bibinfo {author} {\bibfnamefont {A.~J.}\ \bibnamefont
			{Maciejewski}}\ and\ \bibinfo {author} {\bibfnamefont {T.}~\bibnamefont
			{Stachowiak}},\ }\bibfield  {title} {\bibinfo {title} {A novel approach to
			the spectral problem in the two photon {Rabi} model},\ }\href
	{https://doi.org/10.1088/1751-8121/aa6fb8} {\bibfield  {journal} {\bibinfo
			{journal} {J. Phys. A}\ }\textbf {\bibinfo {volume} {50}},\ \bibinfo {pages}
		{244003} (\bibinfo {year} {2017})}\BibitemShut {NoStop}%
	\bibitem [{\citenamefont {Xie}\ \emph {et~al.}(2019)\citenamefont {Xie},
		\citenamefont {Duan},\ and\ \citenamefont {Chen}}]{PhysRevA.99.013809}%
	\BibitemOpen
	\bibfield  {author} {\bibinfo {author} {\bibfnamefont {Y.-F.}\ \bibnamefont
			{Xie}}, \bibinfo {author} {\bibfnamefont {L.}~\bibnamefont {Duan}},\ and\
		\bibinfo {author} {\bibfnamefont {Q.-H.}\ \bibnamefont {Chen}},\ }\bibfield
	{title} {\bibinfo {title} {Generalized quantum {Rabi} model with both one-
			and two-photon terms: A concise analytical study},\ }\href
	{https://doi.org/10.1103/PhysRevA.99.013809} {\bibfield  {journal} {\bibinfo
			{journal} {Phys. Rev. A}\ }\textbf {\bibinfo {volume} {99}},\ \bibinfo
		{pages} {013809} (\bibinfo {year} {2019})}\BibitemShut {NoStop}%
	\bibitem [{\citenamefont {Cong}\ \emph {et~al.}(2019)\citenamefont {Cong},
		\citenamefont {Sun}, \citenamefont {Liu}, \citenamefont {Ying},\ and\
		\citenamefont {Luo}}]{PhysRevA.99.013815}%
	\BibitemOpen
	\bibfield  {author} {\bibinfo {author} {\bibfnamefont {L.}~\bibnamefont
			{Cong}}, \bibinfo {author} {\bibfnamefont {X.-M.}\ \bibnamefont {Sun}},
		\bibinfo {author} {\bibfnamefont {M.}~\bibnamefont {Liu}}, \bibinfo {author}
		{\bibfnamefont {Z.-J.}\ \bibnamefont {Ying}},\ and\ \bibinfo {author}
		{\bibfnamefont {H.-G.}\ \bibnamefont {Luo}},\ }\bibfield  {title} {\bibinfo
		{title} {Polaron picture of the two-photon quantum {Rabi} model},\ }\href
	{https://doi.org/10.1103/PhysRevA.99.013815} {\bibfield  {journal} {\bibinfo
			{journal} {Phys. Rev. A}\ }\textbf {\bibinfo {volume} {99}},\ \bibinfo
		{pages} {013815} (\bibinfo {year} {2019})}\BibitemShut {NoStop}%
	\bibitem [{\citenamefont {{Armenta Rico}}\ \emph {et~al.}(2020)\citenamefont
		{{Armenta Rico}}, \citenamefont {Maldonado-Villamizar},\ and\ \citenamefont
		{Rodriguez-Lara}}]{Rico2020}%
	\BibitemOpen
	\bibfield  {author} {\bibinfo {author} {\bibfnamefont {R.~J.}\ \bibnamefont
			{{Armenta Rico}}}, \bibinfo {author} {\bibfnamefont {F.~H.}\ \bibnamefont
			{Maldonado-Villamizar}},\ and\ \bibinfo {author} {\bibfnamefont {B.~M.}\
			\bibnamefont {Rodriguez-Lara}},\ }\bibfield  {title} {\bibinfo {title}
		{Spectral collapse in the two-photon quantum {R}abi model},\ }\href
	{https://doi.org/10.1103/PhysRevA.101.063825} {\bibfield  {journal} {\bibinfo
			{journal} {Phys. Rev. A}\ }\textbf {\bibinfo {volume} {101}},\ \bibinfo
		{pages} {063825} (\bibinfo {year} {2020})}\BibitemShut {NoStop}%
	\bibitem [{\citenamefont {Zou}\ \emph {et~al.}(2020)\citenamefont {Zou},
		\citenamefont {Zhang}, \citenamefont {Xu}, \citenamefont {Huang},\ and\
		\citenamefont {Liao}}]{zou2019multiphoton}%
	\BibitemOpen
	\bibfield  {author} {\bibinfo {author} {\bibfnamefont {F.}~\bibnamefont
			{Zou}}, \bibinfo {author} {\bibfnamefont {X.-Y.}\ \bibnamefont {Zhang}},
		\bibinfo {author} {\bibfnamefont {X.-W.}\ \bibnamefont {Xu}}, \bibinfo
		{author} {\bibfnamefont {J.-F.}\ \bibnamefont {Huang}},\ and\ \bibinfo
		{author} {\bibfnamefont {J.-Q.}\ \bibnamefont {Liao}},\ }\bibfield  {title}
	{\bibinfo {title} {Multiphoton blockade in the two-photon {Jaynes-Cummings}
			model},\ }\href {https://doi.org/10.1103/PhysRevA.102.053710} {\bibfield
		{journal} {\bibinfo  {journal} {Phys. Rev. A}\ }\textbf {\bibinfo {volume}
			{102}},\ \bibinfo {pages} {053710} (\bibinfo {year} {2020})}\BibitemShut
	{NoStop}%
	\bibitem [{\citenamefont {Garbe}\ \emph {et~al.}(2017)\citenamefont {Garbe},
		\citenamefont {Egusquiza}, \citenamefont {Solano}, \citenamefont {Ciuti},
		\citenamefont {Coudreau}, \citenamefont {Milman},\ and\ \citenamefont
		{Felicetti}}]{garbe_superradiant_2017}%
	\BibitemOpen
	\bibfield  {author} {\bibinfo {author} {\bibfnamefont {L.}~\bibnamefont
			{Garbe}}, \bibinfo {author} {\bibfnamefont {I.~L.}\ \bibnamefont
			{Egusquiza}}, \bibinfo {author} {\bibfnamefont {E.}~\bibnamefont {Solano}},
		\bibinfo {author} {\bibfnamefont {C.}~\bibnamefont {Ciuti}}, \bibinfo
		{author} {\bibfnamefont {T.}~\bibnamefont {Coudreau}}, \bibinfo {author}
		{\bibfnamefont {P.}~\bibnamefont {Milman}},\ and\ \bibinfo {author}
		{\bibfnamefont {S.}~\bibnamefont {Felicetti}},\ }\bibfield  {title} {\bibinfo
		{title} {Superradiant phase transition in the ultrastrong-coupling regime of
			the two-photon {Dicke} model},\ }\href
	{https://doi.org/10.1103/PhysRevA.95.053854} {\bibfield  {journal} {\bibinfo
			{journal} {Phys. Rev. A}\ }\textbf {\bibinfo {volume} {95}},\ \bibinfo
		{pages} {053854} (\bibinfo {year} {2017})}\BibitemShut {NoStop}%
	\bibitem [{\citenamefont {Chen}\ and\ \citenamefont
		{Zhang}(2018)}]{PhysRevA.97.053821}%
	\BibitemOpen
	\bibfield  {author} {\bibinfo {author} {\bibfnamefont {X.-Y.}\ \bibnamefont
			{Chen}}\ and\ \bibinfo {author} {\bibfnamefont {Y.-Y.}\ \bibnamefont
			{Zhang}},\ }\bibfield  {title} {\bibinfo {title} {Finite-size scaling
			analysis in the two-photon {Dicke} model},\ }\href
	{https://doi.org/10.1103/PhysRevA.97.053821} {\bibfield  {journal} {\bibinfo
			{journal} {Phys. Rev. A}\ }\textbf {\bibinfo {volume} {97}},\ \bibinfo
		{pages} {053821} (\bibinfo {year} {2018})}\BibitemShut {NoStop}%
	\bibitem [{\citenamefont {Cui}\ \emph {et~al.}(2019)\citenamefont {Cui},
		\citenamefont {H\'ebert}, \citenamefont {Gr\'emaud}, \citenamefont
		{Rousseau}, \citenamefont {Guo},\ and\ \citenamefont
		{Batrouni}}]{PhysRevA.100.033608}%
	\BibitemOpen
	\bibfield  {author} {\bibinfo {author} {\bibfnamefont {S.}~\bibnamefont
			{Cui}}, \bibinfo {author} {\bibfnamefont {F.}~\bibnamefont {H\'ebert}},
		\bibinfo {author} {\bibfnamefont {B.}~\bibnamefont {Gr\'emaud}}, \bibinfo
		{author} {\bibfnamefont {V.~G.}\ \bibnamefont {Rousseau}}, \bibinfo {author}
		{\bibfnamefont {W.}~\bibnamefont {Guo}},\ and\ \bibinfo {author}
		{\bibfnamefont {G.~G.}\ \bibnamefont {Batrouni}},\ }\bibfield  {title}
	{\bibinfo {title} {Two-photon {Rabi-Hubbard} and {Jaynes-Cummings-Hubbard}
			models: Photon-pair superradiance, {Mott} insulator, and normal phases},\
	}\href {https://doi.org/10.1103/PhysRevA.100.033608} {\bibfield  {journal}
		{\bibinfo  {journal} {Phys. Rev. A}\ }\textbf {\bibinfo {volume} {100}},\
		\bibinfo {pages} {033608} (\bibinfo {year} {2019})}\BibitemShut {NoStop}%
	\bibitem [{\citenamefont {Garbe}\ \emph {et~al.}(2019)\citenamefont {Garbe},
		\citenamefont {Wade}, \citenamefont {Minganti}, \citenamefont {Shammah},
		\citenamefont {Felicetti},\ and\ \citenamefont
		{Nori}}]{garbe2019dissipation}%
	\BibitemOpen
	\bibfield  {author} {\bibinfo {author} {\bibfnamefont {L.}~\bibnamefont
			{Garbe}}, \bibinfo {author} {\bibfnamefont {P.}~\bibnamefont {Wade}},
		\bibinfo {author} {\bibfnamefont {F.}~\bibnamefont {Minganti}}, \bibinfo
		{author} {\bibfnamefont {N.}~\bibnamefont {Shammah}}, \bibinfo {author}
		{\bibfnamefont {S.}~\bibnamefont {Felicetti}},\ and\ \bibinfo {author}
		{\bibfnamefont {F.}~\bibnamefont {Nori}},\ }\bibfield  {title} {\bibinfo
		{title} {Dissipation-induced bistability in the two-photon {Dicke} model},\
	}\href {https://arxiv.org/abs/1911.11694} {\bibfield  {journal} {\bibinfo
			{journal} {arXiv:1911.11694}\ } (\bibinfo {year} {2019})}\BibitemShut
	{NoStop}%
	\bibitem [{\citenamefont {Cui}\ \emph {et~al.}(2020)\citenamefont {Cui},
		\citenamefont {Grémaud}, \citenamefont {Guo},\ and\ \citenamefont
		{Batrouni}}]{Cui2020}%
	\BibitemOpen
	\bibfield  {author} {\bibinfo {author} {\bibfnamefont {S.}~\bibnamefont
			{Cui}}, \bibinfo {author} {\bibfnamefont {B.}~\bibnamefont {Grémaud}},
		\bibinfo {author} {\bibfnamefont {W.}~\bibnamefont {Guo}},\ and\ \bibinfo
		{author} {\bibfnamefont {G.~G.}\ \bibnamefont {Batrouni}},\ }\bibfield
	{title} {\bibinfo {title} {Nonlinear two-photon {Rabi-Hubbard} model:
			{Superradiance}, photon, and photon-pair {Bose-Einstein} condensates},\ }\href
	{https://arxiv.org/abs/2006.09412} {\bibfield  {journal} {\bibinfo  {journal}
			{arXiv:2006.09412}\ } (\bibinfo {year} {2020})}\BibitemShut {NoStop}%
	\bibitem [{\citenamefont {Villas-Boas}\ and\ \citenamefont
		{Rossatto}(2019)}]{PhysRevLett.122.123604}%
	\BibitemOpen
	\bibfield  {author} {\bibinfo {author} {\bibfnamefont {C.~J.}\ \bibnamefont
			{Villas-Boas}}\ and\ \bibinfo {author} {\bibfnamefont {D.~Z.}\ \bibnamefont
			{Rossatto}},\ }\bibfield  {title} {\bibinfo {title} {Multiphoton
			{Jaynes-Cummings} {Model}: {Arbitrary} {Rotations} in {Fock} {Space} and
			{Quantum} {Filters}},\ }\href
	{https://doi.org/10.1103/PhysRevLett.122.123604} {\bibfield  {journal}
		{\bibinfo  {journal} {Phys. Rev. Lett.}\ }\textbf {\bibinfo {volume} {122}},\
		\bibinfo {pages} {123604} (\bibinfo {year} {2019})}\BibitemShut {NoStop}%
	\bibitem [{\citenamefont {Casanova}\ \emph {et~al.}(2018)\citenamefont
		{Casanova}, \citenamefont {Puebla}, \citenamefont {Moya-Cessa},\ and\
		\citenamefont {Plenio}}]{casanova2018connecting}%
	\BibitemOpen
	\bibfield  {author} {\bibinfo {author} {\bibfnamefont {J.}~\bibnamefont
			{Casanova}}, \bibinfo {author} {\bibfnamefont {R.}~\bibnamefont {Puebla}},
		\bibinfo {author} {\bibfnamefont {H.}~\bibnamefont {Moya-Cessa}},\ and\
		\bibinfo {author} {\bibfnamefont {M.~B.}\ \bibnamefont {Plenio}},\ }\bibfield
	{title} {\bibinfo {title} {Connecting \emph{n}th order generalised quantum
			{Rabi} models: Emergence of nonlinear spin-boson coupling via spin
			rotations},\ }\href {https://www.nature.com/articles/s41534-018-0096-9}
	{\bibfield  {journal} {\bibinfo  {journal} {Npj Quantum Inf.}\ }\textbf
		{\bibinfo {volume} {4}},\ \bibinfo {pages} {47} (\bibinfo {year}
		{2018})}\BibitemShut {NoStop}%
	\bibitem [{\citenamefont {Gonz\'alez-Guti\'errez}\ and\ \citenamefont
		{Torres}(2019)}]{PhysRevA.99.023854}%
	\BibitemOpen
	\bibfield  {author} {\bibinfo {author} {\bibfnamefont {C.~A.}\ \bibnamefont
			{Gonz\'alez-Guti\'errez}}\ and\ \bibinfo {author} {\bibfnamefont {J.~M.}\
			\bibnamefont {Torres}},\ }\bibfield  {title} {\bibinfo {title} {Atomic {Bell}
			measurement via two-photon interactions},\ }\href
	{https://doi.org/10.1103/PhysRevA.99.023854} {\bibfield  {journal} {\bibinfo
			{journal} {Phys. Rev. A}\ }\textbf {\bibinfo {volume} {99}},\ \bibinfo
		{pages} {023854} (\bibinfo {year} {2019})}\BibitemShut {NoStop}%
	\bibitem [{\citenamefont {Loudon}(2000)}]{Book_Loudon2000}%
	\BibitemOpen
	\bibfield  {author} {\bibinfo {author} {\bibfnamefont {R.}~\bibnamefont
			{Loudon}},\ }\href {https://books.google.fr/books?id=AEkfajgqldoC} {\emph
		{\bibinfo {title} {The Quantum Theory of Light}}}\ (\bibinfo  {publisher}
	{OUP Oxford},\ \bibinfo {year} {2000})\BibitemShut {NoStop}%
	\bibitem [{\citenamefont {Rumi}\ and\ \citenamefont
		{Perry}(2010)}]{Rumi2010Two-Photon}%
	\BibitemOpen
	\bibfield  {author} {\bibinfo {author} {\bibfnamefont {M.}~\bibnamefont
			{Rumi}}\ and\ \bibinfo {author} {\bibfnamefont {J.~W.}\ \bibnamefont
			{Perry}},\ }\bibfield  {title} {\bibinfo {title} {Two-photon absorption: An
			overview of measurements and principles},\ }\href
	{https://doi.org/10.1364/AOP.2.000451} {\bibfield  {journal} {\bibinfo
			{journal} {Adv. Opt. Photon.}\ }\textbf {\bibinfo {volume} {2}},\ \bibinfo
		{pages} {451} (\bibinfo {year} {2010})}\BibitemShut {NoStop}%
	\bibitem [{\citenamefont {Rivera}\ \emph {et~al.}(2016)\citenamefont {Rivera},
		\citenamefont {Kaminer}, \citenamefont {Zhen}, \citenamefont {Joannopoulos},\
		and\ \citenamefont {Solja{\v c}i{\'c}}}]{Rivera2016Shrinking}%
	\BibitemOpen
	\bibfield  {author} {\bibinfo {author} {\bibfnamefont {N.}~\bibnamefont
			{Rivera}}, \bibinfo {author} {\bibfnamefont {I.}~\bibnamefont {Kaminer}},
		\bibinfo {author} {\bibfnamefont {B.}~\bibnamefont {Zhen}}, \bibinfo {author}
		{\bibfnamefont {J.~D.}\ \bibnamefont {Joannopoulos}},\ and\ \bibinfo {author}
		{\bibfnamefont {M.}~\bibnamefont {Solja{\v c}i{\'c}}},\ }\bibfield  {title}
	{\bibinfo {title} {Shrinking light to allow forbidden transitions on the
			atomic scale},\ }\href {https://doi.org/10.1126/science.aaf6308} {\bibfield
		{journal} {\bibinfo  {journal} {Science}\ }\textbf {\bibinfo {volume}
			{353}},\ \bibinfo {pages} {263} (\bibinfo {year} {2016})}\BibitemShut
	{NoStop}%
	\bibitem [{\citenamefont {Flick}\ \emph {et~al.}(2018)\citenamefont {Flick},
		\citenamefont {Rivera},\ and\ \citenamefont {Narang}}]{Flick2018Strong}%
	\BibitemOpen
	\bibfield  {author} {\bibinfo {author} {\bibfnamefont {J.}~\bibnamefont
			{Flick}}, \bibinfo {author} {\bibfnamefont {N.}~\bibnamefont {Rivera}},\ and\
		\bibinfo {author} {\bibfnamefont {P.}~\bibnamefont {Narang}},\ }\bibfield
	{title} {\bibinfo {title} {Strong light-matter coupling in quantum chemistry
			and quantum photonics},\ }\href
	{https://doi.org/doi:10.1515/nanoph-2018-0067} {\bibfield  {journal}
		{\bibinfo  {journal} {Nanophotonics}\ }\textbf {\bibinfo {volume} {7}},\
		\bibinfo {pages} {1479} (\bibinfo {year} {2018})}\BibitemShut {NoStop}%
	\bibitem [{\citenamefont {Blais}\ \emph {et~al.}(2021)\citenamefont {Blais},
		\citenamefont {Grimsmo}, \citenamefont {Girvin},\ and\ \citenamefont
		{Wallraff}}]{blais2020circuit}%
	\BibitemOpen
	\bibfield  {journal} {  }\bibfield  {author} {\bibinfo {author} {\bibfnamefont
			{A.}~\bibnamefont {Blais}}, \bibinfo {author} {\bibfnamefont {A.~L.}\
			\bibnamefont {Grimsmo}}, \bibinfo {author} {\bibfnamefont {S.~M.}\
			\bibnamefont {Girvin}},\ and\ \bibinfo {author} {\bibfnamefont
			{A.}~\bibnamefont {Wallraff}},\ }\bibfield  {title} {\bibinfo {title}
		{Circuit quantum electrodynamics},\ }\href
	{https://doi.org/10.1103/RevModPhys.93.025005} {\bibfield  {journal}
		{\bibinfo  {journal} {Rev. Mod. Phys.}\ }\textbf {\bibinfo {volume} {93}},\
		\bibinfo {pages} {025005} (\bibinfo {year} {2021})}\BibitemShut {NoStop}%
	\bibitem [{\citenamefont {Breuer}\ and\ \citenamefont
		{Petruccione}(2007)}]{BookBreuer2007}%
	\BibitemOpen
	\bibfield  {author} {\bibinfo {author} {\bibfnamefont {H.-P.}\ \bibnamefont
			{Breuer}}\ and\ \bibinfo {author} {\bibfnamefont {F.}~\bibnamefont
			{Petruccione}},\ }\href@noop {} {\emph {\bibinfo {title} {The Theory of Open
				Quantum Systems}}}\ (\bibinfo  {publisher} {Oxford University Press, New
		York},\ \bibinfo {year} {2007})\BibitemShut {NoStop}%
	\bibitem [{\citenamefont {Rivas}\ \emph {et~al.}(2010)\citenamefont {Rivas},
		\citenamefont {Plato}, \citenamefont {Huelga},\ and\ \citenamefont
		{Plenio}}]{Rivas2010}%
	\BibitemOpen
	\bibfield  {author} {\bibinfo {author} {\bibfnamefont {{\'{A}}.}~\bibnamefont
			{Rivas}}, \bibinfo {author} {\bibfnamefont {A.~D.~K.}\ \bibnamefont {Plato}},
		\bibinfo {author} {\bibfnamefont {S.~F.}\ \bibnamefont {Huelga}},\ and\
		\bibinfo {author} {\bibfnamefont {M.~B.}\ \bibnamefont {Plenio}},\ }\bibfield
	{title} {\bibinfo {title} {Markovian master equations: A critical study},\
	}\href {https://doi.org/10.1088/1367-2630/12/11/113032} {\bibfield  {journal}
		{\bibinfo  {journal} {New J. Phys.}\ }\textbf {\bibinfo {volume} {12}},\
		\bibinfo {pages} {113032} (\bibinfo {year} {2010})}\BibitemShut {NoStop}%
	\bibitem [{\citenamefont {Giorgi}\ \emph {et~al.}(2020)\citenamefont {Giorgi},
		\citenamefont {Saharyan}, \citenamefont {Gu\'erin}, \citenamefont {Sugny},\
		and\ \citenamefont {Bellomo}}]{Giorgi2020}%
	\BibitemOpen
	\bibfield  {author} {\bibinfo {author} {\bibfnamefont {G.~L.}\ \bibnamefont
			{Giorgi}}, \bibinfo {author} {\bibfnamefont {A.}~\bibnamefont {Saharyan}},
		\bibinfo {author} {\bibfnamefont {S.}~\bibnamefont {Gu\'erin}}, \bibinfo
		{author} {\bibfnamefont {D.}~\bibnamefont {Sugny}},\ and\ \bibinfo {author}
		{\bibfnamefont {B.}~\bibnamefont {Bellomo}},\ }\bibfield  {title} {\bibinfo
		{title} {Microscopic and phenomenological models of driven systems in
			structured reservoirs},\ }\href {https://doi.org/10.1103/PhysRevA.101.012122}
	{\bibfield  {journal} {\bibinfo  {journal} {Phys. Rev. A}\ }\textbf {\bibinfo
			{volume} {101}},\ \bibinfo {pages} {012122} (\bibinfo {year}
		{2020})}\BibitemShut {NoStop}%
	\bibitem [{\citenamefont {Streltsov}\ \emph {et~al.}(2017)\citenamefont
		{Streltsov}, \citenamefont {Adesso},\ and\ \citenamefont
		{Plenio}}]{Streltsov2017}%
	\BibitemOpen
	\bibfield  {author} {\bibinfo {author} {\bibfnamefont {A.}~\bibnamefont
			{Streltsov}}, \bibinfo {author} {\bibfnamefont {G.}~\bibnamefont {Adesso}},\
		and\ \bibinfo {author} {\bibfnamefont {M.~B.}\ \bibnamefont {Plenio}},\
	}\bibfield  {title} {\bibinfo {title} {\emph{Colloquium}: Quantum coherence
			as a resource},\ }\href {https://doi.org/10.1103/RevModPhys.89.041003}
	{\bibfield  {journal} {\bibinfo  {journal} {Rev. Mod. Phys.}\ }\textbf
		{\bibinfo {volume} {89}},\ \bibinfo {pages} {041003} (\bibinfo {year}
		{2017})}\BibitemShut {NoStop}%
	\bibitem [{\citenamefont {Wang}\ \emph {et~al.}(2018)\citenamefont {Wang},
		\citenamefont {Wu}, \citenamefont {Cui},\ and\ \citenamefont
		{Wang}}]{Wang2018}%
	\BibitemOpen
	\bibfield  {author} {\bibinfo {author} {\bibfnamefont {Z.}~\bibnamefont
			{Wang}}, \bibinfo {author} {\bibfnamefont {W.}~\bibnamefont {Wu}}, \bibinfo
		{author} {\bibfnamefont {G.}~\bibnamefont {Cui}},\ and\ \bibinfo {author}
		{\bibfnamefont {J.}~\bibnamefont {Wang}},\ }\bibfield  {title} {\bibinfo
		{title} {Coherence enhanced quantum metrology in a nonequilibrium optical
			molecule},\ }\href {https://doi.org/10.1088/1367-2630/aab03a} {\bibfield
		{journal} {\bibinfo  {journal} {New J. Phys.}\ }\textbf {\bibinfo {volume}
			{20}},\ \bibinfo {pages} {033034} (\bibinfo {year} {2018})}\BibitemShut
	{NoStop}%
	\bibitem [{\citenamefont {Smirne}\ \emph {et~al.}(2019)\citenamefont {Smirne},
		\citenamefont {Lemmer}, \citenamefont {Plenio},\ and\ \citenamefont
		{Huelga}}]{Smirne2019}%
	\BibitemOpen
	\bibfield  {author} {\bibinfo {author} {\bibfnamefont {A.}~\bibnamefont
			{Smirne}}, \bibinfo {author} {\bibfnamefont {A.}~\bibnamefont {Lemmer}},
		\bibinfo {author} {\bibfnamefont {M.~B.}\ \bibnamefont {Plenio}},\ and\
		\bibinfo {author} {\bibfnamefont {S.~F.}\ \bibnamefont {Huelga}},\ }\bibfield
	{title} {\bibinfo {title} {Improving the precision of frequency estimation
			via long-time coherences},\ }\href {https://doi.org/10.1088/2058-9565/aaf43d}
	{\bibfield  {journal} {\bibinfo  {journal} {Quantum Sci. Technol.}\ }\textbf
		{\bibinfo {volume} {4}},\ \bibinfo {pages} {025004} (\bibinfo {year}
		{2019})}\BibitemShut {NoStop}%
	\bibitem [{\citenamefont {Orlando}\ \emph {et~al.}(1999)\citenamefont
		{Orlando}, \citenamefont {Mooij}, \citenamefont {Tian}, \citenamefont
		{van~der Wal}, \citenamefont {Levitov}, \citenamefont {Lloyd},\ and\
		\citenamefont {Mazo}}]{Orlando99}%
	\BibitemOpen
	\bibfield  {author} {\bibinfo {author} {\bibfnamefont {T.~P.}\ \bibnamefont
			{Orlando}}, \bibinfo {author} {\bibfnamefont {J.~E.}\ \bibnamefont {Mooij}},
		\bibinfo {author} {\bibfnamefont {L.}~\bibnamefont {Tian}}, \bibinfo {author}
		{\bibfnamefont {C.~H.}\ \bibnamefont {van~der Wal}}, \bibinfo {author}
		{\bibfnamefont {L.~S.}\ \bibnamefont {Levitov}}, \bibinfo {author}
		{\bibfnamefont {S.}~\bibnamefont {Lloyd}},\ and\ \bibinfo {author}
		{\bibfnamefont {J.~J.}\ \bibnamefont {Mazo}},\ }\bibfield  {title} {\bibinfo
		{title} {Superconducting persistent-current qubit},\ }\href
	{https://doi.org/10.1103/PhysRevB.60.15398} {\bibfield  {journal} {\bibinfo
			{journal} {Phys. Rev. B}\ }\textbf {\bibinfo {volume} {60}},\ \bibinfo
		{pages} {15398} (\bibinfo {year} {1999})}\BibitemShut {NoStop}%
	\bibitem [{\citenamefont {van~der Wal}\ \emph {et~al.}(2000)\citenamefont
		{van~der Wal}, \citenamefont {ter Haar}, \citenamefont {Wilhelm},
		\citenamefont {Schouten}, \citenamefont {Harmans}, \citenamefont {Orlando},
		\citenamefont {Lloyd},\ and\ \citenamefont {Mooij}}]{vanderwal00}%
	\BibitemOpen
	\bibfield  {author} {\bibinfo {author} {\bibfnamefont {C.~H.}\ \bibnamefont
			{van~der Wal}}, \bibinfo {author} {\bibfnamefont {A.~C.~J.}\ \bibnamefont
			{ter Haar}}, \bibinfo {author} {\bibfnamefont {F.~K.}\ \bibnamefont
			{Wilhelm}}, \bibinfo {author} {\bibfnamefont {R.~N.}\ \bibnamefont
			{Schouten}}, \bibinfo {author} {\bibfnamefont {C.~J. P.~M.}\ \bibnamefont
			{Harmans}}, \bibinfo {author} {\bibfnamefont {T.~P.}\ \bibnamefont
			{Orlando}}, \bibinfo {author} {\bibfnamefont {S.}~\bibnamefont {Lloyd}},\
		and\ \bibinfo {author} {\bibfnamefont {J.~E.}\ \bibnamefont {Mooij}},\
	}\bibfield  {title} {\bibinfo {title} {Quantum {S}uperposition of
			{M}acroscopic {P}ersistent-{C}urrent {S}tates},\ }\href
	{https://doi.org/10.1126/science.290.5492.773} {\bibfield  {journal}
		{\bibinfo  {journal} {Science}\ }\textbf {\bibinfo {volume} {290}},\ \bibinfo
		{pages} {773} (\bibinfo {year} {2000})}\BibitemShut {NoStop}%
	\bibitem [{\citenamefont {Tinkham}(2004)}]{tinkham2004introduction}%
	\BibitemOpen
	\bibfield  {author} {\bibinfo {author} {\bibfnamefont {M.}~\bibnamefont
			{Tinkham}},\ }\href@noop {} {\emph {\bibinfo {title} {Introduction to
				Superconductivity, {\normalfont 2nd ed.}}}}\ (\bibinfo  {publisher} {Dover
		Publications, New York},\ \bibinfo {year} {2004})\BibitemShut {NoStop}%
	\bibitem [{\citenamefont {Landau}\ and\ \citenamefont
		{Lifshitz}(1976)}]{BookLandau1976Mechanics}%
	\BibitemOpen
	\bibfield  {author} {\bibinfo {author} {\bibfnamefont {L.~D.}\ \bibnamefont
			{Landau}}\ and\ \bibinfo {author} {\bibfnamefont {E.~M.}\ \bibnamefont
			{Lifshitz}},\ }\href@noop {} {\emph {\bibinfo {title} {Course of Theoretical
				Physics, {\normalfont 3rd ed.}}}}\ (\bibinfo  {publisher}
	{Butterworth-Heinemann},\ \bibinfo {year} {1976}), Vol. 1\BibitemShut {NoStop}%
	\bibitem [{Note1()}]{Note1}%
	\BibitemOpen
	\bibinfo {note} {For convenience of the reader, we report in this note the
		formulas for the inductances normal frequencies in terms of the energetic
		circuit parameters: \begin {equation*} \omega _{L}^- = \protect \frac {4
			\protect \sqrt {2}}{\hbar } \protect \sqrt {\protect \frac {E_C \protect
				\tilde {E}_C E_L}{N \protect \tilde {E}_C + 2 \eta E_C}}, \hskip 1em\relax
		\omega _{L}^+ =\protect \frac {4}{\hbar } \protect \sqrt {\protect \frac
			{\protect \tilde {E}_C E_L}{\eta }}. \end {equation*} These formulas are
		equivalent to those of Eq.~\protect \textup {\hbox {\mathsurround \z@
				\protect \normalfont (\ignorespaces \ref {eq:NormalFrequencies}\unskip
				\@@italiccorr )}}.}\BibitemShut {Stop}%
\end{thebibliography}

%

\end{document}